\documentclass[12pt,twoside,a4paper]{article}
\usepackage{amsmath,amssymb,amsthm}

\newtheorem{Thm}{Theorem}[section]
\newtheorem{Lem}[Thm]{Lemma}

\newtheorem{Cor}[Thm]{Corollary}

\theoremstyle{definition}
\newtheorem{Exa}[Thm]{Example}

\theoremstyle{remark}
\newtheorem{Rem}[Thm]{Remark}

\newcommand{\R}{\mathbb{R}}
\newcommand{\Z}{\mathbb{Z}}
\newcommand{\N}{\mathbb{N}}
\newcommand{\Q}{\mathbb{Q}}
\newcommand{\C}{\mathbb{C}}
\newcommand{\U}{\mathbb{U}}

\newcommand{\cA}{\mathcal{A}}

\newcommand{\cD}{\mathcal{D}}
\newcommand{\cE}{\mathcal{E}}

\newcommand{\cH}{\mathcal{H}}
\newcommand{\cL}{\mathcal{L}}

\newcommand{\cU}{\mathcal{U}}

\newcommand{\al}{\alpha}

\newcommand{\ga}{\gamma}
\newcommand{\Ga}{\Gamma}
\newcommand{\de}{\delta}
\newcommand{\De}{\Delta}

\newcommand{\ka}{\kappa}
\newcommand{\la}{\lambda}
\newcommand{\La}{\Lambda}
\newcommand{\om}{\omega}
\newcommand{\Om}{\Omega}
\newcommand{\si}{\sigma}

\renewcommand{\phi}{\varphi}
\newcommand{\te}{\theta}

\newcommand{\ze}{\zeta}

\newcommand{\bd}{\partial}
\newcommand{\set}[2]{\{#1:\,\text{#2}\}}
\newcommand{\sm}{\setminus}
\newcommand{\sub}{\subset}

\newcommand{\ov}{\overline}
\newcommand{\wt}{\widetilde}

\newcommand{\ot}{\otimes}
\newcommand{\lgl}{\langle}
\newcommand{\rgl}{\rangle}

\newcommand{\dir}{\cD}
\newcommand{\nb}{\nabla}
\newcommand{\enn}{E_{\nb}}
\newcommand{\dv}{\bd v}
\newcommand{\dV}{\bd V}

\newcommand{\ds}{\De s}
\newcommand{\hE}{\hat\cE}
\newcommand{\dpw}{\bd_+w}
\newcommand{\dmw}{\bd_-w}

\newcommand{\dnb}{\de\nb}
\newcommand{\dxi}{\de\xi}
\newcommand{\he}{\hat e}
\newcommand{\cjnb}{d_{\nb}^*}
\newcommand{\lan}{\land}
\newcommand{\ane}{\land\hskip 0truemm e}

\newcommand{\form}[1]{\Om_{\dir}^#1(\cA)}
\newcommand{\eform}[1]{\cE\ot\Om_{\dir}^#1(\cA)}
\newcommand{\hform}[1]{\cE\ot\Om_{\dir}^#1(\cA)\ot\hat\cE}

\newcommand{\id}{\operatorname{id}}
\newcommand{\re}{\operatorname{Re}}
\newcommand{\im}{\operatorname{Im}}
\newcommand{\tr}{\operatorname{Tr}}
\newcommand{\sg}{\operatorname{sgn}}
\newcommand{\diag}{\operatorname{diag}}
\newcommand{\dist}{\operatorname{dist}}
\newcommand{\YM}{\operatorname{YM}}
\newcommand{\End}{\operatorname{End}}

\begin{document}

\title{Metrics of nonpositive curvature on graph-manifolds
and electromagnetic fields on graphs}

\author{Sergei Buyalo\footnote{Supported by RFFI Grants
RFFI Grants 96-01-00674 and CRDF Grant RM1-169.}}

\date{}

\maketitle

\begin{abstract}
A 3-dimensional graph-manifold is composed from simple blocks
which are products of compact surfaces with boundary by the
circle. Its global structure may be as complicated as one likes and is
described by a graph which might be an arbitrary graph. A metric of
nonpositive  curvature on such a manifold, if it exists, can be
described essentially by a finite number of parameters which satisfy a
geometrization equation. The aim of the work is to show that this
equation is a discrete version of the Maxwell equations of classical
electrodynamics, and its solutions, i.e., metrics of nonpositive
curvature, are critical configurations of the same sort of action
which describes the interaction of an electromagnetic field with a
scalar charged field. We establish this analogy in the framework
of the spectral calculus (noncommutative geometry) of A. Connes.
\end{abstract}

\section{Introduction}
The aim of this work is to establish a precise analogy between metrics of
nonpositive curvature on 3-dimensional graph-manifolds on the one hand and the
interaction of an electromagnetic field
$\nb$
with a scalar charged field
$\xi$
on the other hand. This analogy is quite unexpected, however, various
geometric effects related to those metrics coincide with effects arising in a
discrete model of the interaction of fields
$\nb$
and
$\xi$.

Metrics of nonpositive sectional curvature on a graph-manifold have a
special structure, and most essential geometric information encoded in a
metric can be described by a finite number of parameters. These parameters
satisfy a compatibility equation, which is similar to the Laplace equation
on graphs. However, there are essential distinctions from the Laplace
equation, and seeking for a continual analog of the compatibility
equation we came to the Euler-Lagrange equations for the action
$$S(\nb,\xi)=\YM(\nb)+E_{\nb}(\xi)-m^2\|\xi\|^2,$$
which describes the interaction of a complex-valued scalar field
$\xi$
and an electromagnetic field
$\nb$,
where
$\YM(\nb)=\|\nb\circ\nb\|^2$
is the Yang-Mills action of
$\nb$, $E_{\nb}(\xi)=\|\nb\xi\|^2$
the energy,
$m$
the mass of
$\xi$.
The field
$\xi$
is a section of the trivial line bundle
$\cE$
over a graph, and
$\nb$
is a connection on
$\cE$.
For a discrete space such as a graph (i.e. a collection of vertices and a
collection of edges between them) the definitions of the connection
$\nb$,
the covariant differential
$\nb\xi$
and the curvature
$\nb\circ\nb$
are possible in the framework of the spectral calculus of A.~Connes developed
in \cite{Con1}, and we use it to establish the mentioned above analogy.

The suggested approach to the discretization problem also
unveil a mechanism of exclusions and degenerations hidden
in the Euler-Lagrange equations, see Sect.~{\bf\ref{sect:excdeg}.}

\medskip
{\bf Acknowledgment.} The author is grateful to L.~Khalfin for the
attention to this work and valuable remarks, V.~Kobel'skii for numerous
discussions and V.~Schroeder for the invitation in the University
of Z\"urich, where this work was finished, and the hospitality.

\section{The compatibility equation}

We consider a closed orientable (and oriented) graph-manifold
$M=\cup_{v\in V}M_v$,
which consists of a finite set
$V$
of building blocks
$M_v=F_v\times S^1$,
where
$F_v$
is a compact surface with boundary, different from the disk and the annulus.
The blocks
$M_v$
are glued along boundary tori
$T_w=(\bd F_v)_w\times S^1$, $w\in\dv$,
where
$\dv$
is the set of the boundary components of
$M_v$,
and the index
$w$
points on such a component. For more details about graph-manifolds see
Appendix~A.

Any metric of nonpositive sectional curvature of the manifold
$M$
locally splits along each block
$M_v$
into the metric product
$ds_F^2+dl^2$,
where
$l$
is the coordinate along the factor
$S^1$,
and
$ds_F^2$
is a metric of nonpositive curvature on a surface (the global splitting along
$M_v$
may not exist and, as the rule, it does not exist). Furthermore, all fibers
$f\times S^1$, $f\in F_v$,
are closed geodesics of the same length
$l_v$,
and the boundary tori can always be chosen to be flat and totally geodesic.
The most essential information about the metric is encoded in collections of
lengths
$\{l_v\}_{v\in V}$
of fibers of blocks
$M_v$
and angles
$\{\om_w\}_{w\in W}$
between the fibers of adjacent blocks
$M_v$, $M_{v'}$, $w=(v,v')$
on the gluing torus
$T_w$,
and, therefore, it is described by a finite set of parameters. The sets
$V$
of blocks and
$W$
of gluing tori are correspondingly the vertex set and the set of oriented
edges of a graph
$\Ga$,
which is called {\it the graph} of the manifold
$M$.
Here a vertex
$v\in V$
is initial for an edge
$w\in W$
iff
$w\in\dv$
(notation:
$v=\dmw$).
We denote by
$\dpw$
the terminal vertex of the edge
$w$.
It is not excluded that
$\dmw=\dpw$,
i.e. different boundary components of
$M_v$
may be glued with each other; in that case the edge
$w$
is a loop in
$\Ga$.
However, we require that for any edge
$w=(v,v')\in W$
the fibers
$S_v^1$, $S_{v'}^1$
of corresponding adjacent blocks
$M_v$, $M_{v'}$
are not homotopic on the gluing torus
$T_w$.

The collections of lengths
$\{l_v\}$
and angles
$\{\om_w\}$
correspond to a metric of nonpositive sectional curvature on
$M$
iff they are a solution to {\it the compatibility equation}
\begin{equation}\label{eq:comp}
k_vl_v-\sum_{w\in\dv}\frac{\cos\om_w}{b_w}l_{\dpw}=0,
     \quad v\in V,
\end{equation}
satisfying the conditions
$l_v>0$
for all
$v\in V$
and
$0<\om_w=\om_{-w}<\pi$
for all
$w\in W$.
Here the coefficients
$k_v\in\Q$, $b_w=b_{-w}\in\N$
are topological invariants of the manifold
$M$
(see Appendix~A). In this case we say that the collections
$\{l_v\}$, $\{\om_w\}$
define {\it an isometric state} of the system~(\ref{eq:comp}).

\begin{Exa}[Dipole]\label{Exa:dipole} The graph
$\Ga$
consists of two vertices
$v_0$, $v_1$
connected by an edge. The compatibility equation has the form

\parbox{10cm}{
\begin{eqnarray*}\label{eqn:dipole}
 k_0l_0-\frac{\cos\om}{b}l_1&=&0\\
-\frac{\cos\om}{b}l_0+k_1l_1&=&0.
\end{eqnarray*}
}\hfill
\parbox{1cm}{\begin{eqnarray}\end{eqnarray}}

\noindent
Isometric states exist in the next cases.

(i) $k_0=k_1=0$.
Then
$\om=\pi/2$,
and the ratio
$l_0/l_1$
may be arbitrary;

(ii) $0<k_0k_1b^2<1$.
Then
$\cos^2\om=k_0k_1b^2$, $\sg(\cos\om)=\sg(k_j)$
and
$$l_0^2/l_1^2=k_1/k_0.$$
In what follows this example plays a key role.
\end{Exa}

\begin{Exa}[Monopole]\label{Exa:monopole} The graph
$\Ga$
is a loop with vertex
$v$.
The compatibility equation~(\ref{eq:comp}) has the form
\begin{equation}\label{eqn:monopole}
(k-\frac{2}{b}\cos\om)l=0
\end{equation}
for the corresponding coefficients
$k$, $b$.
Isometric states exist iff
$|k|b<2$.
In that case
$\cos\om=kb/2$
and
$l>0$
is arbitrary.
\end{Exa}

\subsection{Comparison with the Laplace equation on graphs}

The Laplacian
$\De:L^2(V)\to L^2(V)$
on a graph
$\Ga=\Ga(V,W)$
is defined as
$$\De f(v)=f(v)-\frac{1}{|\dv|}\sum_{w\in\dv}f(\dpw),$$
where
$L^2(V)$
is the (real) Hilbert space with the scalar product
$$\lgl f, f'\rgl=\sum_{v\in V}|\dv|f(v)f'(v).$$
Then
$\De=d^*d$,
the differential
$d:L^2(V)\to L_{\text{odd}}^2(W)$
is
$df(w)=f(\dpw)-f(\dmw)$,
and
$L_{\text{odd}}^2(W)$
is the Hilbert space of odd functions on
$W$
with the scalar product
$$\lgl\phi,\psi\rgl=\frac{1}{2}\sum_{w\in W}\phi(w)\psi(w).$$
Solutions to the Laplace equation
$\De f=0$
minimize the action
$E(f)=\|df\|^2$.

The compatibility equation~(\ref{eq:comp}) looks like
the Laplace equation with "variable coefficients". However, the factors
$\cos\om_w$
change properties of the solutions and indicate the hidden presence
of a connection
$\nb$,
an electromagnetic field: it will be clear, that the usual differential
$d$
has to be replaced by a covariant differential
$\nb$.

Recall that an electromagnetic field is described as a connection
$\nb$
on the principal (trivial) bundle
$M^4\times\U(1)\to M^4$;
a scalar charged field is a section
$\xi:M^4\to M^4\times\C$
of the bundle
$M^4\times\C\to M^4$,
where
$M^4$
is the Minkowski space. Their interaction minimizes the action
$$S(\nb,\xi)=\|F\|^2+\|\nb\xi\|^2-m^2\|\xi\|^2,$$
where
$F=\nb\circ\nb$
is the curvature of
$\nb$.
In other words, the connection
$\nb$
and the section
$\xi$
satisfy to the Euler-Lagrange equation for the action
$S$

\begin{align*}
  d^*F+\xi^*\nb\xi-\xi(\nb\xi)^*&=0& \tag{$+$} \\
  (\nb^*\nb-m^2)\xi&=0.& \tag{$++$}
\end{align*}
One should add to these the Bianchi's identity
$dF=0$,
which gives (after a choice of coordinates in
$M^4$)
the first pair of the Maxwell equations
$$\text{div {\bf H}}=0,\qquad \text{rot {\bf E}}=-\bd_t{\text{\bf H}}.$$
The equation (+) gives the second pair
$$\text{div {\bf E}}=J_0,\qquad \text{rot {\bf H}}=
  \bd_t\text{{\bf E}}+\text{{\bf J}},$$
and (++) is the wave equation.

Roughly speaking, a metric of nonpositive curvature on a graph-manifold
$M$
may be interpreted as a scalar charged field
$\xi:V\to V\times\C$
on the vertex set of the graph
$\Ga$
of
$M$
interacting with a connection
$\nb$
on the bundle
$V\times\C\to V$.
The connection
$\nb$
may be interpreted as a collection of conformal structures on gluing tori;
the field
$\xi$
contains the information about the lengths of fibers; all together they
minimize the action
$$S(\nb,\xi)=\|F\|^2+\|\nb\xi\|^2-m^2\|\xi\|^2.$$
Therefore, the isometric states of the equation~(\ref{eq:comp})
are extremals of the action
$S$,
and the equation~(\ref{eq:comp}) itself turns out to be an analog of the wave
equation~(++).

\section{Geometry of a two-point space}\label{sect:gts}

The program described at the end of the previous section can be realized in
the framework of the spectral calculus of A.~Connes. Here we collect necessary
facts of that formalism restricting to the case of finite dimensional
algebras. Furtheremore, all notions are illustrated by an example of the
dipole, the simplest nontrivial example, to which the formalism is
applicable (this justifies the title of the section). At the same time, this
example turns out to be fundamental for us, because the case of general graphs
can be reduced to the case of dipole using the decomposition principle (see
Sect.~{\bf\ref{sect:decomp}}). Detailed discussion of the spectral calculus
can be found in
\cite{Con1}--\cite{Con3}, \cite{V-Gr}, see also \cite{KPPW},
\cite{Sch-Z}, \cite{C-M}, \cite{Ka}, \cite{K-W}, \cite{Sit}.

\subsection{Space}

The role of a space is played by an involutive algebra
$\cA$.
For the dipole
$V=\{v_0, v_1\}$
the algebra
$\cA$
is the function algebra
$\{V\to\C\}=\C^2$
with involution
$a=(a_0,a_1)\mapsto a^*=(\ov a_0,\ov a_1)$.

\subsection{Riemannian metric and the spectral differential}

A geometry on
$V$
is specified by a representation of
$\cA$
is a Hilbert space
$\cH$,
``the tangent bundle'',
and a selfadjoint compact operator of unit length
$ds\in L(\cH)$.
{\it The spectral triple}
$\{\cA, \cH, ds\}$
plays the role of a Riemannian manifold. In the case
of the dipole the algebra
$\cA$
is represented in the Hilbert space
$\cH=\C^2$
with the scalar product
$\lgl a,b\rgl=a_0\ov b_0+a_1\ov b_1$
by multiplication operators
$$a=(a_0,a_1)\mapsto\pi(a)=\diag(a_0,a_1):\C^2\to\C^2.$$
The unit length operator is defined as

$$ds=\left[\begin{array}{cc}
              0&-i\ds\\
              i\ds&0\end{array}\right],\quad\ds\in\R.$$
The inverse operator
$\dir=ds^{-1}$
(the Dirac operator)
allows to differentiate functions
$a\in\cA$:
the spectral differential is defined as the operator
$$da:=i[\dir,\pi(a)]\in L(\cH).$$
The operator
$da$
is an element of the space
$\form 1\subset L(\cH)$
of 1-forms, which consists of the operators
$$\om=\sum_ja_0^jda_1^j=i\sum_ja_0^j[\dir,a_1^j],\quad a_i^j\in\cA$$
(for brevity, we identify
$a$
and
$\pi(a)$).
Obviously,
$(da)^*=da^*$.

For
$a=(a_0,a_1)$
we have
\[da=\left[\begin{array}{cc}
             0&\frac{1}{\ds}(a_1-a_0)\\
       \frac{1}{\ds}(a_1-a_0)&0
           \end{array}\right].\]
The metric on
$V$
is defined by the Connes' formula
\begin{equation}\label{eq:conndist}
\dist(v_0,v_1)=\sup\set{\left|a(v_0)-a(v_1)\right|}
   {$a\in\cA,\ \|da\|\le 1$},
\end{equation}
which gives
$\dist(v_0,v_1)=|\ds|$.

\subsection{The line bundle and gauge transformations}

The trivial line bundle
$V\times\C\to V$
is completely characterized by its sections which form the trivial right
$\cA$-module
$\cE$.
The algebra
$\cA$
acts on
$\cE$
from the right,
$(\xi,a)\mapsto\xi a\in\cE$
for
$\xi\in\cE$, $a\in\cA$.
Though the module
$\cE$
is isomorphic to the algebra
$\cA$
(considered as
$\cA$-module)
there is no canonical isomorphism. An isomorphism
$\cE\to\cA$
is defined by the choice of a basis
$e\in\cE$.
The group of gauge transformations
$\cU=\{u\in\cA\,|\,u^*u=uu^*=1\}$
acts on
$\cE$
as
$(\xi,u)\mapsto\xi u^*$, $\xi\in\cE$.
We consider
$\cU$
as the set of distinguished bases
$e$
of the bundle
$\cE$:
the coordinate of a section
$\xi\in\cE$
is the function
$a=a(\xi,e)\in\cA$,
defined by the relation
$\xi=ea$.
When changing the basis
$e\mapsto e'=eu^*$
the coordinate transforms as follows
$a(\xi,e')=ua(\xi,e)$.

In the case of the dipole the group
$\cU$
consists of the elements of form
$u=(e^{i\al_0},e^{i\al_1})$, $\al_0$, $\al_1\in\R$.
For
$\xi\in\cE$
and a basis
$e\in\cU$
we write
$a(\xi,e)=(\xi_0,\xi_1)$.

\subsection{Connections on
$\cE$}

A connection on
$\cE$
is a linear mapping
$\nb:\cE\to\eform 1$
satisfying the Leibniz rule
$$\nb(\xi a)=\nb\xi\cdot a+\xi\ot da,$$
$\xi\in\cE$, $a\in\cA$.
For a basis
$e\in\cU$
a connection
$\nb$
is given by the vector potential
$\Phi\in\form 1$ defined by the relation
$$\nb e=e\ot\Phi.$$
Therefore, for
$\xi=ea\in\cE$
we have
$$\nb\xi=\nb e\cdot a+e\ot da=e\ot(da+\Phi a).$$
The gauge transformation group
$\cU$
acts on vector potentials of the connection
$\nb$
as follows. For
$e'=eu^*$, $\nb e'=e'\ot\Phi'$
we have
$$\Phi'=udu^*+u\Phi u^*.$$
Indeed, on the one hand, we have
$\nb e'=e'\ot\Phi'=eu^*\ot\Phi'=e\ot u^*\Phi'$,
on the other hand,
$\nb e'=\nb(eu^*)=\nb e\cdot u^*+e\ot du^*=e\ot(du^*+\Phi u^*)$.

In the case of a dipole representing
$\Phi$
as an operator
\[\Phi=i\left[\begin{array}{cc}
              0&\phi_{01}\ds\\
             \phi_{10}\ds&0
             \end{array}\right],\quad\phi_{01},\phi_{10}\in\C,\]
for
$\xi=(\xi_0,\xi_1)\in\cE$
we have
\[\nb\xi=e\ot\left[\begin{array}{cc}
     0&\frac{1}{\ds}(\xi_1-\xi_0)+i\phi_{01}\ds\xi_1\\
     \frac{1}{\ds}(\xi_1-\xi_0)+i\phi_{10}\ds\xi_0&0
     \end{array}\right].\]

\subsection{The space of
$k$-forms $\form k$.}

The space of
$k$-forms
$\form k\subset L(\cH)$
consists of the operators of form
\begin{equation}
\om=\sum_ja_0^jda_1^j\dots da_k^j,
\end{equation}
$a_i^j\in\cA$.
The exterior differential
$d:\form k\to\form {{k+1}}$
can be computed by the formula
\begin{equation}\label{eq:difform}
d\om=\sum_jda_0^j\dots da_k^j
\end{equation}
(it is well defined for the dipole case because if
$\sum_ja_0^jda_1^j\dots da_k^j=0$,
then
$\sum_jda_0^j da_1^j\dots a_k^j=0$,
which follows from the fact that the operator
$ds^2$
commutes with the algebra
$\cA$).

Using
$da\cdot b=d(ab)-adb$
for
$a$, $b\in\cA$
we have
$\si\om\in\form{{l+k}}$
for
$\si\in\form l$, $\om\in\form k$.
Furtheremore, the Leibniz rule holds
$$d(\si\om)=d\si\cdot\om+(-1)^l\si d\om.$$

\subsection{The curvature of a connection
$\nb$}

A connection
$\nb:\cE\to\eform 1$
extends to a differentiation
$\nb:\eform k\to\eform{{k+1}}$
by the formula
$$\nb(e\ot\om):=\nb e\cdot\om+e\ot d\om$$
for
$\om\in\form k$,
where
$e\in\cU$
is a basis of
$\cE$.
For
$\si\in\form l$
we have
\begin{eqnarray*}
    \nb(e\ot\om\cdot\si)&=&
    \nb e\cdot\om\cdot\si+e\ot d\om\cdot\si+(-1)^ke\ot\om\cdot d\si\\
   &=&\left(\nb(e\ot\om)\right)\si+(-1)^ke\ot\om\cdot d\si.
\end{eqnarray*}

{\it The curvature} of a connetion
$\nb$
is defined as the operator
$$\nb^2=\nb\circ\nb:\cE\to\eform 2.$$
With respect to a basis
$e\in\cU$,
we have
$\nb^2e=e\ot\te$,
where
$\te=d\Phi+\Phi^2\in\form 2$
is {\it the curvature operator}.
Indeed,
$$\nb^2e=\nb(e\ot\Phi)=\nb e\cdot\Phi+e\ot d\Phi
        =e\ot(d\Phi+\Phi^2).$$

\begin{Lem}\label{Lem:curvtensor} The curvature
$\nb^2$
is an
$\cA$-linear
operator, i.e.
$$\nb^2(\xi\cdot a)=\nb^2\xi\cdot a$$
for each
$\xi\in\cE$
and
$a\in\cA$.
\end{Lem}

\begin{proof} Let
$e\in\cU$
be a basis of the module
$\cE$. It suffices to show that
$\nb^2(ea)=\nb^2e\cdot a$
for
$a\in\cA$.
We have
\begin{eqnarray*}
  \nb^2(ea)&=&\nb\left(e\ot(da+\Phi a)\right)\\
   &=&e\ot\Phi(da+\Phi a)+e\ot d(da+\Phi a)\\
   &=&e\ot(\Phi da+\Phi^2 a+d^2a+d\Phi a-\Phi da)\\
   &=&e\ot(d\Phi+\Phi^2)a.
\end{eqnarray*}
\end{proof}

This implies the gauge invariance of the curvature. Let
$\te'$
be the curvature operator of a connection
$\nb$
with respect to a basis
$e'=eu^*$, $\nb^2e'=e'\ot\te'$.
Then
$$\te'=u \te u^*.$$
Indeed, we have
$e'\ot\te'=e\ot u^*\te'$.
On the other hand,
$\nb^2e'=\nb^2e\cdot u^*$
by Lemma~\ref{Lem:curvtensor}. Hence
$u^*\te'=\te u^*.$

To compute the curvature operator
$\te$
in the case of the dipole, one introduces the projectors
$p=(1,0)$, $1-p=(0,1)\in\cA$.
Then
$$\Phi=-i\phi_{01}\ds^2pdp-i\phi_{10}\ds^2(1-p)dp.$$
Using that
$(dp)^2=\ds^{-2}\cdot 1$,
one obtains
$d\Phi=-i(\phi_{01}-\phi_{10})\cdot 1$.
Further,
$\Phi^2=-\phi_{01}\phi_{10}\ds^2\cdot 1$.
Thus
$$\te=d\Phi+\Phi^2
   =i(\phi_{10}-\phi_{01}+i\phi_{01}\phi_{10}\ds^2)\cdot1.$$
In particular, we see that for an unitary connection
$\nb$,
i.e.
$\Phi^*+\Phi=0$, or $\ov\phi_{01}=\phi_{10}$
the norm of the curvature operator
$$\|\te\|^2=\left(\left|\frac{1}{\ds}+i\phi_{01}\ds\right|^2-
            \frac{1}{\ds^2}\right)^2$$
achieves its minimum along the circle
$U_{\ds}\subset\C$
of radius
$1/\ds^2$
centered at
$z_{\ds}=i/\ds^2$
which goes to the real line
$\R=\{\im\phi_{01}=0\}$
as
$\ds\to 0$.
The gauge transformation group
$\cU$
acts transitively on
$U_{\ds}$.

\subsection{The Bianchi identity}

Let
$\te$
be the curvature operator of a connection
$\nb$, $\te=d\Phi+\Phi^2$.
Then {\it the Bianchi identity} takes place
$$d\te=[\te,\Phi].$$
Indeed,
$d\te=d^2\Phi+d\Phi\cdot\Phi-\Phi\cdot d\Phi$.
Using
$d^2=0$
and
$d\Phi=\te-\Phi^2$
we obtain
$$d\te=(\te-\Phi^2)\Phi-\Phi(\te-\Phi^2)=\te\Phi-\Phi\te.$$
In the case of the dipole, the operator
$\te$
is a scalar operator. Thus the both parts of the Bianchi identity are equal
to zero,
$$d\te=0=[\te,\Phi].$$

\subsection{Hermitian structures on
$\cE$}
Let
$\La:\cE\to\cE$
the multiplication operator by a positive element of the algebra
$\cA$.
In the coordinates with respect to a basis
$e\in\cU$,
the operator
$\La$
is defined by
$$a(\La\xi,e)=\la a(\xi,e),$$
where
$\xi\in\cE$, $\la=\la(\La,e)\in\cA$
is the matrix of the map
$\La$
with respect to
$e$, $\la^*=\la$.
For
$e'=eu^*$
we have
$\la'=u\la u^*$,
where
$\la'=\la(\La,e')$.

{\it A hermitian structure} on
$\cE$
plays the role of a fiberwise hermitian metric.
The standard hermitian structure is defined by
$$(\xi,\eta)=b^\ast a\in\cA$$
for
$\xi=ea$, $\eta=eb\in\cE$.
This definition is independent of the choice of a basis
$e\in\cU$.
In general case, a hermitian structure is given by
$$(\xi,\eta)_\La=(\La\xi,\eta)\in\cA.$$
Therefore, we have
$$(\xi,\eta)=b^\ast\la a$$
with respect to a basis
$e$,
where
$\xi=ea$, $\eta=eb$.
In the case of the dipole a hermitian structure is given by positive numbers
$\la_0$, $\la_1$
and
$(\xi,\eta)_{\La}(v_0)=\ov b_0\la_0a_0$,
$(\xi,\eta)_{\La}(v_1)=\ov b_1\la_1a_1$
for
$a=(a_0,a_1)$, $b=(b_0,b_1)$.

\subsection{Connections compatible with a hermitian structure}

On says that a connection
$\nb$
on
$\cE$
is {\it compatible} with a hermitian structure
$(\ ,\ )_{\La}$
or it is
$\La$-hermitian,
if
\begin{equation*}
d(\xi,\eta)_{\La}=(\nb\xi,\eta)_{\La}+(\xi,\nb\eta)_{\La}\tag{$\dag$}
\end{equation*}
for all
$\xi$, $\eta\in\cE$.
Here both sides of the equality are 1-forms, i.e. elements of
$\form 1$.
For
$\xi=ea$, $\eta=eb$
we have
$$(\nb\xi,\eta)_{\La}=\left(e\ot(da+\Phi a),\eta\right)_{\La}
    :=(e,\eta)_{\La}(da+\Phi a)$$
and
$$(\xi,\nb\eta)_{\La}=\left(\xi,e\ot(db+\Phi b)\right)_{\La}
    :=(db+\Phi b)^*(\xi,e)_{\La}.$$

\begin{Lem}\label{Lem:compcon} A connection
$\nb$
is compatible with a hermitian structure
$\La$
iff
with respect to some (and hence any) basis one has
\begin{equation*}
d\la=\la\Phi+\Phi^*\la,\tag{$\ddag$}
\end{equation*}
where
$\Phi$
is the vector potential of the connection
$\nb$
with respect to
$e$, $\la=\la(\La,e)$.
\end{Lem}

\begin{proof} For
$\xi=ea$, $\eta=eb\in\cE$
we have
\begin{eqnarray*}
  d(\xi,\eta)_{\La}&=&d(b^*\la a)
         =db^*\la a+b^*d\la a+b^*\la da\\
  (\nb\xi,\eta)_{\La}&=&(e,\eta)_{\La}(da+\Phi a)
         =b^*\la da+b^*\la\Phi a\\
  (\xi,\nb\eta)_{\La}&=&(db+\Phi b)^*(\xi,e)_{\La}
         =db^*\la a+b^*\Phi^*\la a,
\end{eqnarray*}
where we used
$(db)^*=db^*$.
Thus ($\dag$) is equivalent to ($\ddag$).
\end{proof}

\begin{Rem} The relation~($\ddag$) is the unique one from the relations
above which the author did not find in the literature on the spectral calculus
accessible to him. However, the using of general hermitial structures on
$\cE$
and the formula~($\ddag$) plays an important role in what follows.
\end{Rem}

For the case of the dipole a simple calculation shows that
($\ddag$) is equivalent to
\begin{equation}\label{eq:compcon}
\la_0(1+i\phi_{01}\ds^2)=\la_1(1+i\ov\phi_{10}\ds^2),
\end{equation}
where
$\la=(\la_0,\la_1)$,
$\Phi=-i\phi_{01}\ds^2pdp-i\phi_{10}\ds^2(1-p)dp$.

\begin{Rem} The vector potential
$A$
of an electromagnetic field
$\nb$
is a 1-form on the Minkowski space
$M^4$
with values in the Lie algebra
$u(1)=i\R$
of
$\U(1)$,
thus
$A^*+A=0$.
It follows that for any hermitian structure
$\la$
on
$\cE=\{M^4\to M^4\times\C\}$
we have
$d\la=\la A+A^*\la=0$,
i.e. the function
$\la$
is constant. One can interpret this as the conservation law of the electric
charge.
\end{Rem}

\subsection{Hermitian metrics}

For a hermitian structure
$\La$
on the bundle
$\cE$
the product
$(\xi,\eta)_{\La}\in\cA$
is an operator in
$\cH$.
The corresponding {\it hermitian metric} on
$\cE$
is obtained by integration of the fiberwise hermitian product
$$\lgl\xi,\eta\rgl=\tr(\xi,\eta)_{\La}\in\C$$
for
$\xi$, $\eta\in\cE$.
In the case of the dipole taking a coordinates
$\xi=ea$, $\eta=eb$, $\la=\la(\La,e)=(\la_0,\la_1)$,
we obtain
$$\lgl\xi,\eta\rgl=\tr(b^*\la a)=
   \ov\eta_0\la_0\xi_0+\ov\eta_1\la_1\xi_1.$$

Any hermatian metric on
$\cE$
is canonically extended to hermitian metrics on
$\cA$-modules
$\eform k$, $\hat\cE=\{\cE\to\cA\}$, $\hform k$
(the tensor products are taken over
$\cA$).
This makes possible to define the norms of operators
$\nb\xi\in\eform 1$, $\nb^2\in\hform 2$
(for more details see Appendix~B)
and to introduce an action
$S=S(\nb,\xi)$.

\subsection{An interaction Lagrangian}

A connection
$\nb$
on
$\cE$
compatible with a hermitian structure
$(\ ,\ )_{\La}$
can be considered as analog of an electromagnetic field, and the sections
$\xi\in\cE$
describe scalar charged particles. The energy of such a field
$\xi$
in the presence of
$\nb$
is defined as
$\enn(\xi)=\lgl\nb\xi,\nb\xi\rgl$.

To compute
$\enn(\xi)$
in the case of the dipole, we introduce a basis
$e\in\cU$
and obtain
$\xi=ea=(\xi_0,\xi_1)$, $\nb\xi=e\ot\si$, $\si=da+\Phi a\in\form 1$,
where
\[\si=\left[\begin{array}{cc}
        0&\frac{1}{\ds}(\xi_1-\xi_0)+i\phi_{01}\ds\xi_1\\
        \frac{1}{\ds}(\xi_1-\xi_0)+i\phi_{10}\ds\xi_0&0
        \end{array}\right].\]
Thus for
$\la=(\la_0,\la_1)$
we have
\begin{eqnarray*}
   \enn(\xi)&=&\lgl e\ot\si,e\ot\si\rgl=\lgl\si,\la\si\rgl\\
  &=&\la_0\left|\frac{1}{\ds}(\xi_1-\xi_0)+i\phi_{01}\ds\xi_1\right|^2
   +\la_1\left|\frac{1}{\ds}(\xi_1-\xi_0)+i\phi_{10}\ds\xi_0\right|^2.
\end{eqnarray*}

The Yang-Mills action of
$\nb$
is
$\YM(\nb)=\lgl\nb^2,\nb^2\rgl$,
where the curvature
$\nb^2$
is considered as an element of the space
$\hform 2$
equipped with the
$\La$-hermitian
metric. For the dipole case we have
$$\YM(\nb)=\lgl\te\la,\la\te\rgl=(\la_0^2+\la_1^2)
       \left|\phi_{10}-\phi_{01}+i\phi_{01}\phi_{10}\ds^2\right|^2.$$

The interaction of
$\xi$
and
$\nb$
minimizes the action
$$S(\nb,\xi)=\YM(\nb)+\enn(\xi)-m^2\langle\xi,\xi\rangle.$$

For the dipole case we finally obtain
\begin{eqnarray}\label{eqn:action}
    S(\nb,\xi)&=&(\la_0^2+\la_1^2)
  \left|\phi_{10}-\phi_{01}+i\phi_{01}\phi_{10}\ds^2\right|^2\notag\\
  &+&\la_0\left|\frac{1}{\ds}(\xi_1-\xi_0)+i\phi_{01}\ds\xi_1\right|^2
 +\la_1\left|\frac{1}{\ds}(\xi_1-\xi_0)+i\phi_{10}\ds\xi_0\right|^2\notag\\
    &-&m^2\left(\la_0|\xi_0|^2+\la_1|\xi_1|^2\right),
\end{eqnarray}
and the condition~(\ref{eq:compcon}) for
$\nb$
to be
$\La$-hermitian
is fulfilled. Here
$(\xi_0,\xi_1)$
are the coordinates of
$\xi$
with respect to a basis
$e$, $(\la_0,\la_1)$
coefficients of a hermitian structure
$\La$, $\Phi=-i\phi_{01}\ds^2pdp-i\phi_{10}\ds^2(1-p)dp$
the vector potential of
$\nb$.
The action
$S$
is defined in an invariant way independently on the choice of
$e\in\cU$,
thus the expression~(\ref{eqn:action}) is gauge invariant.

\section{The Euler-Lagrange equation for
$S$}\label{sect:eulagr}

The Euler-Lagrange equations for the action
$S$
have the form
\begin{eqnarray}
         \cjnb\nb^2&=&-J\label{eqn:2ndpair}\\
     \nb^*\nb\xi&=&m^2\xi,\label{eqn:wave}
\end{eqnarray}
where
$J=\nb\xi\ot\hat\xi-\xi\ot\hat\nb\hat\xi$
is the current, and they correspond to the usual equations of classical
electrodynamics: (\ref{eqn:2ndpair}) is the analog of the second pair
of the Maxwell equations~(+), (\ref{eqn:wave}) the analog of the
wave equation~(++). In the case of the dipole, the action
$S$
is given by (\ref{eqn:action}), and the Euler-Lagrange
equations have the form
\begin{equation}\label{eq:twopoint2nd}
  \frac{1}{\ds^2}(\psi_{01}\psi_{10}-1)(\la_0\psi_{10}+\la_1\ov\psi_{01})
  =\ov\xi_0\xi_1-\frac{1}{2}\left(\psi_{10}|\xi_0|^2
                             +\ov\psi_{01}|\xi_1|^2\right)
\end{equation}

\parbox{10cm}{
\begin{eqnarray*}\label{eqn:twopointmaxwell}
\frac{1}{ 2}\left(\la_0(\ov\psi_{01}\ov\psi_{10}+1)-\la_0\rho^2\right)\xi_0
  -\la_0\psi_{01}\xi_1&=&0\\
  -\la_1\psi_{10}\xi_0
+\frac{1}{2}\left(\la_1(\psi_{01}\psi_{10}+1)-\la_1\rho^2\right)\xi_1&=&0,
\end{eqnarray*}
}\hfill
\parbox{1cm}{\begin{eqnarray}\end{eqnarray}}

\noindent
where we used the notations
$\psi_{01}=1+i\phi_{01}\ds^2$,
$\psi_{10}=1-i\phi_{10}\ds^2$, $\rho^2=m^2\ds^2$.
We are keeping the coefficients
$\la_0$, $\la_1$
in the equations~(\ref{eqn:twopointmaxwell}) to stress the analogy with the compatibility
equation~(\ref{eq:comp}) in view of the
$\La$-hermitian
condition~(\ref{eq:compcon}), which takes the form
\begin{equation}\label{eq:newcompcon}
\la_0\psi_{01}=\la_1\ov\psi_{10}.
\end{equation}

We show how to obtain the
equations~(\ref{eqn:2ndpair})--(\ref{eqn:twopointmaxwell}) in Appendix~B.
Here we analize the equations~(\ref{eq:twopoint2nd}),
(\ref{eqn:twopointmaxwell})
comparing them with the compatibility equation for the case
of the elementary graphs: dipole~(\ref{eqn:dipole}) and
monopole~(\ref{eqn:monopole}). The general case is considered in
Sect.~{\bf\ref{sect:decomp}.}

Representing
$\xi_0=|\xi_0|e^{i\al_0}$, $\xi_1=|\xi_1|e^{i\al_1}$
we interpret the numbers
$|\xi_0|$, $|\xi_1|$
as proportional to the fiber lengths
$l_0$, $l_1$
of blocks
$M_{v_0}$, $M_{v_1}$
(see Introduction).
Since these lengths are positive, we are interested only in nondegenerate
solutions to the Euler-Lagrange equations, i.e. for which
$\xi_0$, $\xi_1\neq 0$.
The necessary condition for that is vanishing of the determinant of the
system~(\ref{eqn:twopointmaxwell}),
\begin{equation}\label{eq:nondegenerate}
  \left((\ov\psi_{01}\ov\psi_{10}+1)-\rho^2\right)
  \left((\psi_{01}\psi_{10}+1)-\rho^2\right)
  -4\psi_{01}\psi_{10}=0.
\end{equation}

In view of (\ref{eq:newcompcon}) we can assume that
$\psi_{01}=|\psi_{01}|e^{i\ga}$,
$\psi_{10}=|\psi_{10}|e^{-i\ga}$,
and we put
$$\psi:=\la_0|\psi_{01}|=\la_1|\psi_{10}|.$$
Then the condition (\ref{eq:nondegenerate}) has the form
$$\left(\psi^2-(\rho^2-1)\la_0\la_1\right)^2-4\la_0\la_1\psi^2=0$$
or
$$\left(\psi^2-(\rho^2+1)\la_0\la_1\right)^2=4\la_0^2\la_1^2\rho^2,$$
hence
\begin{equation}\label{eq:newnondgr}
   \psi^2=\la_0\la_1(\rho\pm 1)^2.
\end{equation}
It follows from the imaginary part of the
equations~(\ref{eq:twopoint2nd}) -- (\ref{eqn:twopointmaxwell})
that
\begin{equation}\label{eq:impart}
  \al_1-\al_0+\ga=n\pi,\quad n\in\Z,\
\end{equation}
and we obtain from those equations the next (real) system
\begin{equation}\label{eq:realtwopoint2nd}
  \frac{1}{\De s^2}\left(\frac{\psi^2}{\la_0\la_1}-1\right)
  \left(\frac{\la_0}{\la_1}+\frac{\la_1}{\la_0}\right)\psi
  =(-1)^n|\xi_0||\xi_1|-\frac{1}{2}\left(\frac{|\xi_0|^2}{\la_1}
  +\frac{|\xi_1|^2}{\la_0}\right)\psi,
\end{equation}

\parbox{10cm}{
\begin{eqnarray*}\label{eqn:realtwopointmaxwell}
 \frac{1}{2}\left(\frac{\psi^2}{\la_1}+\la_0(1-\rho^2)\right)|\xi_0|
  +(-1)^{n+1}\psi|\xi_1|&=&0\\
   (-1)^{n+1}\psi|\xi_0|
+\frac{1}{2}\left(\frac{\psi^2}{\la_0}+\la_1(1-\rho^2)\right)|\xi_1|&=&0.
\end{eqnarray*}
}\hfill
\parbox{1cm}{\begin{eqnarray}\end{eqnarray}}

\subsection{The case of a massless field
$\xi$}

We consider, first, the case
$m=0$
for the mass
$m$
and, consequently,
$\rho=0$.
Then
$\psi^2=\la_0\la_1$
by (\ref{eq:newnondgr}),
and for a nonzero solution
$\left(|\xi_0|,|\xi_1|\right)$
of (\ref{eqn:realtwopointmaxwell}) we have
$$\la_0^{1/2}|\xi_0|=(-1)^n\la_1^{1/2}|\xi_1|.$$
Thus
$n$
is even and
\begin{equation}\label{eq:mass0sol}
 \frac{|\xi_0|^2}{|\xi_1|^2}=\frac{\la_1}{\la_0}
\end{equation}
cp. Example~\ref{Exa:dipole}(ii). Equation~(\ref{eq:realtwopoint2nd})
is fulfilled automatically, its left and right hand side both vanish.

Therefore, the solutions to the Euler-Lagrange
equation~(\ref{eqn:2ndpair}) -- (\ref{eqn:wave}) for
the action~(\ref{eqn:action}) in the case of a massless field
$\xi$
are as follows
$$\al_1-\al_0+\ga=2n\pi;\quad\psi^2=\la_0\la_1;\quad
   |\xi_0|^2/|\xi_1|^2=\la_1/\la_0.$$
These data define a flat connection
$\nb$
and a parallel field
$\xi$, $\nb\xi=0$,
i.e.
$S(\nb,\xi)=0$.

\subsubsection{Comparing of the spectral and the geometric models}

To interpret a solution
$(l_0, l_1, \om)$
to the compatibility equation~(\ref{eqn:dipole}),
defining a nonpositively curved metric on a graph-manifold
as a field configuration
$(\nb,\xi)$
extremal for the action
$S$,
we put
$$l_0=|\xi_0|,\ l_1=|\xi_1|,\ \la_0=|k_0|,\ \la_1=|k_1|,
 \ k_0|\psi_{01}|=\cos\om/b,\ k_1|\psi_{01}|=\cos\om/b$$
assuming that
$k_0k_1>0$.
These data define a
$\La$-hermitian
connection
$\nb$
on
$\cE$
and a field
$\xi$,
related by the gauge invariance condition
$\al_1-\al_0+\ga=2n\pi$.
It is this choice that lead to a necessity to consider nonconstant hermitian
structures
$\la$
on
$\cE$
and to use the relation~($\dag$). Now the nondegeneracy condition
$\psi^2=\la_0\la_1$
is equvalent to the condition
$\cos^2\om=k_0k_1b^2$
and (\ref{eq:mass0sol}) is exactly the condition
$l_0^2/l_1^2=k_1/k_0$
for the solutions~(ii) of (\ref{eqn:dipole}).
Under this choice, the equation~(\ref{eqn:realtwopointmaxwell})
is the same as the equation~(\ref{eqn:dipole}), if the condition
$\psi^2=\la_0\la_1$
is fulfilled, which in turn ensures that the connection
$\nb$
is flat,
$\YM(\nb)=0$.

The decomposition principle (see Sect.~{\bf\ref{sect:decomp}}) makes
possible to extend this equivalence of the spectral and geometric
models to the case of arbitrary (finite) graphs, and in that sense
a nonpositively curved metric on a graph-manifold can be interpreted
as a field configuration
$(\nb,\xi)$
extremal for the action
$S$,
where
$\nb$
is a flat
$\La$-hermitian connection on
$\cE$
and
$\xi\in\cE$
is a covariantly constant field,
$\nb\xi=0$,
or, in physical terminology, as the interaction of an electromagnetic field
with a massless scalar charged field on the graph of the manifold.

However, this interpretation, even being attractive, is not complete. Its
drawback is first of all the necessity of the condition
$k_0k_1>0$
since otherwise the spectral model disappears,
$\la_0=\la_1=0$.
On the other hand, on a graph-manifold often exist nonpositively curved
metrics for which it is necessary to consider the case
$k_0=k_1=0$, $\om=\pi/2$
by the decomposition principle. Such metrics cannot be interpreted
in terms of massless fields. Another disadvantage of such interpretation
is that the equations~(\ref{eq:realtwopoint2nd}) --
(\ref{eqn:realtwopointmaxwell}) for
$\rho=0$
do not reflect the obvious geometric prohibition
$\om\neq 0$.
The condition
$\om=0$
implies a collaps of the metric, because it means that the fibers of the
adjacent blocks being homotopically distinct closed geodesics on the flat
gluing torus
$T_w$
coincide. These problems will be solved in the next section, where we consider
the fields
$\xi$
with nonzero masses, due to a remarkable mechanism of exclusions and
degenerations hidden in equations~(\ref{eq:realtwopoint2nd}) --
(\ref{eqn:realtwopointmaxwell}).

\section{The mechanism of exclusions and degenerations}\label{sect:excdeg}

Here we describe a mechanism which allows to interpret in terms of the
Euler-Lagrange equations (\ref{eq:realtwopoint2nd}) --
(\ref{eqn:realtwopointmaxwell}) for the action
$S=S(\nb,\xi)$,
(\ref{eqn:action}), degenerations
$\om=\pi/2$ and $\om=0$
for metrics on nonpositive curvature on a graph-manifold. It requires to
consider fields with nonzero masses. Then the
equations~(\ref{eq:realtwopoint2nd}) --
(\ref{eqn:realtwopointmaxwell})
amazingly unveil a geometric information about the interpreted
metric on the graph-manifold, which is absent in the compatibility
equation~(\ref{eq:comp}). Namely, the equation~(\ref{eq:comp})
being linear and homogeneous in lengths does not fix the lengths
of fibers only defining its ratios. However, if
$m\neq 0$,
then the equations~(\ref{eq:realtwopoint2nd}) --
(\ref{eqn:realtwopointmaxwell}) define the values
$|\xi_0|$, $|\xi_1|$,
and for that there is a natural geometric interpretation.

The exclusions and degenerations mechanism is described during the proof of
the following theorem. We put
$\rho=m\ds$
do not assuming that
$\rho$
is positive. Recall that
$m$
is the mass of a field
$\xi$
and
$|\ds|$
is interpreted as the distance between the vertices
$v_0$, $v_1$.

\begin{Thm}\label{Thm:massgap} The Euler-Lagrange
equations~(\ref{eq:realtwopoint2nd}), (\ref{eqn:realtwopointmaxwell})
for the action
$S$
given by (\ref{eqn:action}) possess nondegenerate solutions (i.e.
$|\xi_0||\xi_1|\neq0$) if and only if
$\rho=0$, $1<|\rho|<2$.
For
$|\rho|=1,2$
any solution degenerates, and for the remaining
$\rho\in\R$
there is no solution.
\end{Thm}

\begin{proof} Since the case
$\rho=0$
is already considered, we assume that
$\rho\neq 0$.
Furthemore, we assume that
$\rho>0$,
the case
$\rho<0$
is treated similarly. For a solution
$(|\xi_0|, |\xi_1|)$
to (\ref{eqn:realtwopointmaxwell}) by nondegeneracy
condition~(\ref{eq:newnondgr}) we have
\begin{equation}\label{eq:massgap1}
  \la_0(1\pm\rho)|\xi_0|=(-1)^n\psi|\xi_1|.
\end{equation}
We first show that in the case
$\psi^2=\la_0\la_1(\rho+1)^2$
for the condition~(\ref{eq:newnondgr}) our system possesses
no solution. Indeed,
$\psi=\la_0^{1/2}\la_1^{1/2}(\rho+1)$,
and it follows from (\ref{eq:massgap1}) that
$n$
is even and
$|\xi_0|^2/|\xi_1|^2=\la_1/\la_0$.
Then the left hand side of (\ref{eq:realtwopoint2nd}) is positive,
$$\frac{1}{\ds^2}\rho(\rho+1)(\rho+2)
  \frac{\la_0^2+\la_1^2}{\la_0^{1/2}\la_1^{1/2}}>0,$$
whereas the right hand side is
$$(-1)^n|\xi_0||\xi_1|
   -\frac{1}{2}\left(\frac{|\xi_0|^2}{\la_1}
    +\frac{|\xi_1|^2}{\la_0}\right)\psi
   =-\la_0^{1/2}\la_1^{-1/2}\rho|\xi_0|^2\le 0.$$
Thus we suppose in the sequel that
$\psi^2=\la_0\la_1(\rho-1)^2$.

\noindent
{\bf (a) The case
$0<\rho<1$.} Then
$\psi=\la_0^{1/2}\la_1^{1/2}(1-\rho)$
and it follows from (\ref{eq:massgap1}) that
$\la_0^{1/2}|\xi_0|=(-1)^n\la_1^{1/2}|\xi_1|$.
Thus
$n$
is even and
$|\xi_0|^2/|\xi_1|^2=\la_1/\la_0$.
This time the left hand side of (\ref{eq:realtwopoint2nd}) is negative,
$$\frac{1}{\ds^2}\rho(\rho-2)(1-\rho)
  \frac{\la_0^2+\la_1^2}{\la_0^{1/2}\la_1^{1/2}}<0,$$
whereas the right hand side is
$$(-1)^n|\xi_0||\xi_1|
   -\frac{1}{2}\left(\frac{|\xi_0|^2}{\la_1}
   +\frac{|\xi_1|^2}{\la_0}\right)\psi
   =\la_0^{1/2}\la_1^{-1/2}\rho|\xi_0|^2\ge 0.$$

\noindent
{\bf (b) The case
$\rho=1$.} Then
$\psi=0$
and the system~(\ref{eqn:realtwopointmaxwell}) put no restriction on
$|\xi_0|$, $|\xi_1|$.
The equation~(\ref{eq:realtwopoint2nd}) is reduced to the condition
$|\xi_0||\xi_1|=0$,
which gives a degenerate critical configuration
$(\nb,\xi)$
for the action
$S$.

We assume now that
$\rho>1$. Then
$\psi=\la_0^{1/2}\la_1^{1/2}(\rho-1)$,
and it follows from (\ref{eq:massgap1}) that
$n$
is odd and
$|\xi_0|^2/|\xi_1|^2=\la_1/\la_0$.
The left hand side of (\ref{eq:realtwopoint2nd}) is
$$\frac{1}{\ds^2}\left(\frac{\psi^2}{\la_0\la_1}-1\right)
  \left(\frac{\la_0}{\la_1}+\frac{\la_1}{\la_0}\right)\psi
  =\frac{1}{\ds^2}\rho(\rho-1)(\rho-2)
   \frac{\la_0^2+\la_1^2}{\la_0^{1/2}\la_1^{1/2}},$$
and the right hand side is
$$(-1)^n|\xi_0||\xi_1|
   -\frac{1}{2}\left(\frac{|\xi_0|^2}{\la_1}
   +\frac{|\xi_1|^2}{\la_0}\right)\psi
   =-\la_0^{1/2}\la_1^{-1/2}\rho|\xi_0|^2.$$
Thus for
$\rho>2$
there is no critical configuration
$(\nb,\xi)$
for
$S$,
and for
$\rho=2$
the unique critical configuration is a degenerate one,
$|\xi_0|=|\xi_1|=0$, $\psi=\la_0\la_1$.

It remains to consider

\noindent
{\bf (c) The case
$1<\rho<2$.}
Then equations (\ref{eq:realtwopoint2nd}),
(\ref{eqn:realtwopointmaxwell}) give a critical configuration
$(\nb,\xi)$
for which
$$\al_1-\al_0+\ga=(2n+1)\pi,\quad\psi=\la_0^{1/2}\la_1^{1/2}(\rho-1)$$
and
$$\la_0|\xi_0|^2=\la_1|\xi_1|^2
      =\frac{(\la_0^2+\la_1^2)(\rho-1)(2-\rho)}{\ds^2}.$$
Obviously, we have
$|\xi_0|^2/|\xi_1|^2=\la_1/\la_0$
and the connection
$\nb$
is not flat. This completes the proof of the theorem.
\end{proof}

\subsection{Interpretation of metrics as fields with nonzero masses}

Interpreting a solution
$(l_0,l_1,\om)$
to compatibility equation~(\ref{eqn:dipole}) with
$k_0$, $k_1>0$
(the case
$k_0$, $k_1<0$
is treated similarly)
as a field configuration
$(\nb,\xi)$
extremal for the action
$S$
with a field
$\xi$
of nonzero mass
$m$,
we assume by Theorem~\ref{Thm:massgap} that
$1<\rho<2$
(in the case
$k_0$, $k_1<0$
one should take
$-2<\rho<-1$)
and put
$$\la_0|\xi_0|^2=k_0l_0^2,\ \la_1|\xi_1|^2=k_1l_1^2,\ \psi=\cos\om/b,
  \ \la_0(\rho-1)=k_0,\ \la_1(\rho-1)=k_1.$$
These data define a
$\La$-hermitian
connection
$\nb$
on
$\cE$
and a field
$\xi$
of nonzero mass
$m$, $\rho=m\ds$
related by the gauge invariance condition
$\al_1-\al_0+\ga=(2n+1)\pi$.
Now the nondegeneracy condition
$\psi^2=\la_0\la_1(\rho-1)^2$
is equvalent to the condition
$\cos^2\om=k_0k_1b^2$
and the relation
$|\xi_0|^2/|\xi_1|^2=\la_1/\la_0$
is exactly the condition
$l_0^2/l_1^2=k_1/k_0$.
But now, in contrast to the case
$m=0$,
these data define the lengths
$l_0$, $l_1$,

\begin{equation}\label{eq:lenght}
 \la_0l_0^2=\la_1l_1^2=\frac{(\la_0^2+\la_1^2)(2-\rho)}{\ds^2}
\end{equation}
making the configuration
$(\nb,\xi)$
extremal for the action
$S$.
This raises the question how to understand these equalities in
geometric terms? The equation~(\ref{eq:comp}) is homogeneous
in lengths of fibers, and this reflects the possibility of
scaling the metric by homotheties.

To understand (\ref{eq:lenght}) correctly we will consider
special nonpositively curved metrics on a graph-manifold
$M$,
whose restrictions on each block
$M_v$
are geometric structures modelled on
$H^2\times\R$,
i.e. the metrics
$ds_F^2$
in the local decompositions
$ds_{M_v}^2=ds_F^2+dl^2$
are metrics of the constant curvature
$-1$.
We call such metrics {\it
geometrizations} of
$M$
(see \cite{BK1}--\cite{BK3}).

The knowledge of the lengths
$l_0$, $l_1$
of fibers of adjacent blocks and of the angle
$\om$
between them uniquely defines a flat metric on the gluing torus
$T_w$
and hence it fixes the length of a closed geodesic
$z_w$
on
$T_w$
representing the boundary component
$(\bd F_v)_w$
of the surface
$F_w$
from the decomposition
$M_v=F_v\times S^1$
(the choice of
$z_w$
is in general not unique and it is equivalent to the choice of a Waldhausen
basis, see Appendix~A; however, it does not depend on the metric and one can
assume a Waldhausen basis to be fixed).

In turn, there is a metric of constant curvature
$-1$
on the surface
$F_v$
with geodesic boundary for which the length of each component
$z_w$, $w\in\dv$
is a given positive number. But then the lengths
$l_0$, $l_1$
are uniquely defined by the length of the geodesic
$z_w$,
if a conformal structure on the torus
$T_w$
given by the ratio
$l_0/l_1$
and the angle
$\om$
is fixed.

Namely this property is reflected by (\ref{eq:lenght}). The length of
$z_w$
is proportional to
$1/|\ds|$,
and this is compatible with interpretation
$|\ds|$
as a ``distance'' between blocks
$M_{v_0}$, $M_{v_1}$:
metrics collapse
as
$|\ds|\to\infty$,
and become infinitely ``thick''
as
$|\ds|\to 0$.
In other words, the parameter
$m$
governs the shape of geometrizations of the manifold
$M$
when
$\rho=m\ds$
is fixed.

\subsubsection{Degeneration
$\rho=1$}

Another fact confirming that the spectral model with nonzero mass is adequate
to the geometric model is its possibility to interpret the degeneration
$k_0=k_1=0$,
when
$\om=\pi/2$
and there is no restriction on the fiber lengths
$l_0$, $l_1$
(recall that such interpretation is impossible by massless fields). This case
naturally corresponds to the case
$\rho=1$,
when
$\psi=0$
and
$|\xi_0||\xi_1|=0$.
Then one can take
$\la_0$, $\la_1$
as arbitrary positive numbers, and we obtain a critical configuration
$(\nb,\xi)$
defined by the metric in question. Furthermore, any limit transition
$k_0$, $k_1\to 0+$
can be interpreted by a corresponding limit transition
$\rho\to 1+$.

An adequate description of a much more strong degeneration
$\om\to 0$
of metrics is possible for more general graph-manifolds, whose graph is
different from the dipole. The point is that if a graph-manifold with the
dipole graph admits a nonpositively curved metric, then
$k_0=k_1=0$
and
$\om=\pi/2$
(see Appendix~A),
which corresponds to the case
$\rho=1$.

\subsubsection{Monopole}

We briefly describe the spectral model for the graph
$\Ga$
which is a loop with vertex
$v$,
see Example~\ref{Exa:dipole}(\ref{eqn:monopole}). Here
$\cA=\C$, $\cH=\C^2$,
the representation of
$\cA$
in
$\cH$
is given by matrices
$\pi(a)=\diag(a,a)$.
The rest data are obiously obtained from the already considered dipole case:
$\phi_{01}=\phi_{10}$, $\la_0=\la_1(=\la)$, $\xi_0=\xi_1(=\xi)$
etc. The Euler-Lagrange equation for the action
$S$
possess the same mechanism of exclusions and degenerations as in the dipole
case. We leave the details to the reader. Notice only that
$k_0=k_1=k/2$
by the decomposition principle and the correspondence between the geometric
and the spectral models is given by
$$\la|\xi|=kl,\ \psi=\cos\om/b,\ \la(\rho-1)=k/2$$
(we assume that
$k>0$).
Then for a critical configuration we have
$$\la^2(\rho-1)^2=\psi^2=\cos^2\om/b^2.$$

\section{The decomposition principle}\label{sect:decomp}

The decomposition principle was found by V.~Kobel'skii in 1994
and widely used in works
\cite{BK1}--\cite{BK3}.
It is based on the following simple topological property of the
graph-manifolds (see Appendix~A). Let
$M_v$
be a block of a graph-manifold
$M$,
which is a toral sum
$M'+M''$
of graph-manifolds
$M'$, $M''$
along blocks
$M_{v'}'\subset M'$, $M_{v''}''\subset M''$,
where the vertex
$v$
is the conjunction of the vertices
$v'$, $v''$.
Then we have
$k_v=k_{v'}+k_{v''}$
for the corresponding invariants (charges) of the manifolds
$M$, $M'$, $M''$.
The decomposition principle is related to the graphs of graph-manifolds, more
precisely, to {\it the labeled graphs} being equipped with charges
$k_v$, $v\in V$
and intersection indices
$b_w$, $w\in W$,
which are the coefficients of compatibility
equation~(\ref{eq:comp}). To simplify
the statement, we restrict to the case when the graph
$\Ga$
of a manifold
$M$
is {\it the triplet}, i.e. consists of two edges
$w_0\cup w_1$
with the common vertex
$\dpw_0=v_1=\dmw_1$.
We use notations
$v_0=\dmw_0$, $v_2=\dpw_1$
and
$k_0$, $k_1$, $k_2$
for the corresponding charges,
$b_0$, $b_1$
for the corresponding intersection indices. Any decomposition
$k_1=k_1^++k_1^-$
represents the labeled graph
$\Ga$
as the conjunction of the labeled graphs
$\Ga_0=w_0$, $\Ga_1=w_1$
along the vertex
$v_1$,
where
$\Ga_0$
is equipped with the charges
$k_0$, $k_1^+$
and the intersection index
$b_0$,
and
$\Ga_1$
with
$k_1^-$, $k_2$, $b_1$
correspondingly.

\medskip
\noindent
{\bf The decomposition principle} asserts that

\medskip
\noindent
{\bf(a)} if the graphs
$\Ga_0$, $\Ga_1$
possess isometric states
$(l_0,l_1^+;\om_0)$
and
$(l_1^-,l_2;\om_1)$
correspondingly and
$l_1^+=l_1^-=:l_1$,
then
$(l_0,l_1,l_2;\om_0,\om_1)$
is an isometric state of the graph
$\Ga$;

\noindent
{\bf(b)} conversely, any isometric state
$(l_0,l_1,l_2;\om_0,\om_1)$
of
$\Ga$
canonically defines the decomposition
$\Ga=\Ga_0+\Ga_1$
with
$k_1=k_1^++k_1^-$,
where the numbers
$k_1^+$, $k_1^-$
are defined by the relations
$l_0^2/l_1^2=k_1^+/k_0$, $l_1^2/l_2^2=k_2/k_1^-$
(if
$k_0$
or
$k_1$
vanishes, then
$k_1^+=0$
or correspondingly
$k_1^-=0)$.

In other words, this principle corresponds to a representation of
system~(\ref{eq:comp})
for the graph
$\Ga$

\[\begin{array}{cccc}
  k_0l_0 &-\frac{\cos\om_0}{b_0}l_1 & &=0\\
  &&&\\
  -\frac{\cos\om_0}{b_0}l_0&+k_1l_1&-\frac{\cos\om_1}{b_1}l_2&=0\\
  &&&\\
  &-\frac{\cos\om_1}{b_1}l_1&+k_2l_2&=0\\
\end{array}\]

as the system for two dipoles
$\Ga_0$, $\Ga_1$

\[\begin{array}{cccc}
  k_0l_0 &-\frac{\cos\om_0}{b_0}l_1^+ & &=0\\
  &&&\\
  -\frac{\cos\om_0}{b_0}l_0&+k_1^+l_1^+&&=0\\
  &&&\\
  &k_1^-l_1^- & -\frac{\cos\om_1}{b_1}l_2&=0\\
  &&&\\
  &-\frac{\cos\om_1}{b_1}l_1^-&+k_2l_2&=0\\
\end{array}\]
with additional conditions
$k_1=k_1^++k_1^-$
and
$l_1^+=l_1=l_1^-$.

We stress that a decomposition of labeled graphs
$\Ga=\Ga_0+\Ga_1$
in general does not exist on the level of graph-manifolds, i.e. the
corresponding representation
$M=M_0+M_1$
as a toral sum may not exist and as the rule it does not exist.

This principle is obviously extended to any graph-manifold
$M$
allowing to decompose its labeled graph into elementary labeled graphs --
monopoles and dipoles. Of course, it is meaningful only for
$M$
admitting metrics of nonpositive curvature. Roughly speaking, this principle
describes a parametrization of the set of nonpositively curved metrics
(geometrizations) on
$M$.

\subsection{Spectral model}

Here we describe the spectral model of a geometrization of an arbitrary
graph-manifold
$M$.
As the motivation we use the above decomposition principle.

Let
$\Ga=\Ga(V,W)$
be the graph of
$M$
with the vertex set
$V$
and the set of oriented edges
$W$.
The algebra
$\cA$
(``the space'') is defined as the algebra of functions
$v\mapsto\{\dv\to\C\}$
on the set
$V$,
where
$\dv\subset W$
is the set of edges
$w$
with the common initial vertex
$\dmw=v$.
As ``the tangent bundle'' we take the Hilbert space of sections
$\eta:W\to W\times\C$
of the trivial line bundle
$W\times\C\to W$
with the scalar product
$$\lgl\eta,\eta'\rgl=\sum_{w\in W}\ov\eta'(w)\eta(w).$$
The representation
$\pi$
of
$\cA$
in
$\cH$
is given by
$$\left(\pi(a)\eta\right)_v(w)=a_v(w)\eta(w)$$
for
$w\in\dv$.
The unit length operator
$ds$
is defined as
$$ds(\eta)(w)=i\ds(w)\cdot\eta(-w),$$
where
$\ds:W\to\R\setminus 0$
is an odd function,
$\ds(-w)=-\ds(w)$.

These data define a spectral triple
$\{\cA,\cH,ds\}$
associated with the graph
$\Ga$.
In the dipole case this definition coincides with one given in
Sect.~{\bf\ref{sect:gts}}.
The~Connes' formula~(\ref{eq:conndist}) applied to edges defines
an intrinsic metric on
$\Ga$,
which is called {\it spectral}. Different edges
$w$, $w'$
between the same vertices may have different lengths. It is not difficult to
see that for any intrinsic metric on
$\Ga$
there is an odd function
$\ds:W\to\R\setminus 0$,
for which the lengths of any edge with respect to the given metric and the
corresponding spectral metric coincide. This fact may serve as a starting point
for approximation by (discrete) spectral models of classical ones.

The space of scalar charged fields
$\cE$,
the gauge transformation group
$\cU$,
a hermitian structure
$\La$,
a
$\La$-hermitian connection
$\nb$
on
$\cE$,
its curvature
$\nb\circ\nb$
and the hermitian metrics on
$\cA$-modules
$\cE$, $\eform 1$, $\hform 2$
are described in Sect.~{\bf\ref{sect:gts}}. Hence
we can define the action
$$S(\nb,\xi)=\YM(\nb)+\enn(\xi)-m^2\|\si\|^2$$
and derive its Euler-Lagrange equation~(\ref{eqn:2ndpair}),
(\ref{eqn:wave}).

\subsubsection{Reduction of the gauge group}

The model described above is, obviously, the direct sum of (almost
independent) spectral models for the dipoles
$\Ga_w$, $w\in W$. Only relations between the last are established via the
function
$\ds$
defining the spectral metric on the graph
$\Ga$.
The function
$\ds$
is not a dynamical variable, hence the Euler-Lagrange equation for
$S$
breaks down into a system of
$|W|/2$
independent pairs of equations~(\ref{eq:twopoint2nd}),
(\ref{eqn:twopointmaxwell}), parametrized by (nonoriented)
edges
$(w,-w)$, $w\in W$.
The gauge group
$\cU$
is the group of maps
$\{\dV\to\U(1)\}$,
where
$\dV=\cup_{v\in V}\dv$.
To make the situation less trivial and adopted for approximations,
we restrict
$\cU$
to its subgroup
$\cU_0=\{V\to\U(1)\}$.
It means that for a fixed basis
$e\in\cU_0$
and for every
$\xi=(\xi_w)\in\cE$, $v\in V$
we have
$\arg(\xi_w)=\arg(\xi_{w'})$
for any
$w$, $w'\in\dv$.

Much more essential relations between the dipoles
$\Ga_w$
arise when we fix a correspondence between the spectral and geometrical
models.

\subsubsection{Correspondence between the spectral and geometric models}

To simplify statements we assume that the graph
$\Ga$
of
$M$
has no loops. Fix a metric of nonpositive curvature
(geometrization) on
$M$
(the necessary and sufficient conditions for the existence of such metrics are
established in
\cite{BK2}).
By the decomposition principle it defines the decomposition of the labeled
graph
$\{\Ga, B, K\}$
into the dipoles
$\{\Ga_w, b_w, K_w\}$, $w\in W$,
where
$K=\{k_v\,|\,v\in V\}$
is the collection of charges of the manifold
$M$, $B=\{b_w\,|\,w\in W\}$
are the intersection indices,
$b_w=b_{-w}\in\N$, $K_w=\{k_w,k_{-w}\}$
charges of the dipole
$\Ga_w$, $k_w:=k_{\dmw}$.
Furthemore, the condition
$$k_v=\sum_{w\in\dv}k_w$$
holds for each vertex
$v\in V$.

The geometrization defines for each dipole
$\Ga_w$
its isometric state
$$(l_w,l_{-w};\om_w),$$
where
$l_w:=l_{\dmw}$,
see Example~\ref{Exa:dipole}.
We assume for simplicity that all these isometric states
are nondegenerate, i.e.

\begin{equation}\label{eq:cosnondeg}
 0<\cos^2\om_w=k_wk_{-w}b_w^2<1.
\end{equation}
The case of negative charges
$k_w$, $k_{-w}$
in general cannot be excluded. To describe such cases we pick an orientation
$w=w_s$
of every nonoriented edge
$(w,-w)$
such that
$\sg(\ds(w))=\sg(k_w)=\sg(k_{-w})$.
We let
$W_s\subset W$
be the set of the choosen oriented edges. Furthemore, if the graph
$\Ga$
is not simply connected, then for every its nontrivial circuit
$Z\subset\Ga$
the following balance condition is fulfilled. Assume that
$Z$
is oriented and the orientations
$w=w_Z$
of its edges are compatible with the orientation of
$Z$
(the orientation
$w_Z$
nothing to do with the orientation
$w_s$
above). Then

\begin{equation}\label{eq:prod}
 \prod_{w\in Z}\frac{k_w}{k_{-w}}=1.
\end{equation}
This relation, obviously, follows from the equality

\begin{equation}\label{eq:mainratio}
 \frac{l_{-w}^2}{l_w^2}=\frac{k_w}{k_{-w}}
\end{equation}
for each edge
$w\in W$.

Assuming for simplicity that for any edge
$w\in W_s$
the number
$\rho_w=m\ds(w)$
satisfies the condition
$1<\rho_w<2$,
we put as in Sect.~\ref{sect:excdeg}

\begin{equation}\label{eq:interpr1}
 \la_w|\xi_w|^2=k_wl_w^2;
\end{equation}
\begin{equation}\label{eq:interpr2}
 \la_w(\rho_w-1)=k_w;
\end{equation}
\begin{equation}\label{eq:interpr3}
\psi_w=\cos\om_w/b_w;
\end{equation}
\begin{equation}\label{eq:interpr4}
 \arg(\xi_w)-\arg(\xi_{-w})+\ga_w=(2n+1)\pi,
\end{equation}
where
$\la_w:=\la_{\dmw}$.
For a fixed basis
$e\in\cU_0$
these conditions define a field
$\xi\in\cE$
with coordinates
$\xi_w:=\xi_{\dmw}\in\C$,
a hermitian structure
$\La=(\la_w)$
on
$\cE$
and a
$\La$-hermitian
connection
$\nb$
on
$\cE$,
whose coefficients are defined via parameters
$\psi_w$,$\ga_w$
as in Sect.~{\bf\ref{sect:eulagr}}. Condition~(\ref{eq:interpr4})
of the gauge invariance ensures that the configuration
$(\nb,\xi)$
is well defined, i.e. does not depend on the choice of
$e\in\cU_0$.
It follows from (\ref{eq:cosnondeg}), (\ref{eq:interpr2}) and
(\ref{eq:interpr3}) that the nondegeneracy condition
$$\psi_w^2=\la_w\la_{-w}(\rho_w-1)^2$$
holds, see~(\ref{eq:newnondgr}), and from
(\ref{eq:mainratio}) and (\ref{eq:interpr1}) that
$|\xi_{-w}|^2/|\xi_w|^2=\la_w/\la_{-w}$,
i.e. the field
$\xi$
satisfies the wave equation~(\ref{eqn:wave}).

\subsubsection{Solution of the Maxwell equations}

The field configuration
$(\nb,\xi)$
defined above is critical for the action
$S$,
if it satisfies the equation~(\ref{eqn:2ndpair}).
It means that for any edge
$w\in W_s$
the condition

\begin{equation}\label{eq:lengthdef}
 \la_w|\xi_w|^2=\la_{-w}|\xi_{-w}|^2=\frac{(\la_w^2+\la_{-w}^2)
   (\rho_w-1)(2-\rho_w)}{\ds^2(w)}
\end{equation}
has to be satisfied, see proof of Theorem~\ref{Thm:massgap}(c).
These conditions for different edges
$w$, $w'\in\dv\cap W_s$
are not independent, because for the fiber lengths we have the equality
$l_{\dmw}=l_v=l_{\dmw'}$.
Thus condition~(\ref{eq:lengthdef}) means that there is some relation
between parameters
$\rho_w$, $\rho_{w'}$.
In other words, the field configuration
$(\nb,\xi)$
defined by the geometrization
$(l_w,\om_w)$, $w\in W$
of the manifold
$M$
by (\ref{eq:interpr1}) -- (\ref{eq:interpr4}) is critical for the action
$S$,
if it is possible to satisfy the mentioned relation between
$\rho_w$, $\rho_{w'}$
for all pairs
$w$, $w'\in\dv\cap W_s$, $v\in V$
without a contradiction.

\begin{Thm}\label{Thm:solmaxw} For any geometrization
$(l_w,\om_w)$, $w\in W$
of the manifold
$M$
there exists an (odd) function
$\ds:W\to\R\sm 0$
such that the field configuration
$(\nb,\xi)$
defined by (\ref{eq:interpr1}) -- (\ref{eq:interpr4})
is critical for the action
$S$.
\end{Thm}

\begin{proof} If there is an edge
$w$
with
$k_w=0$,
then
$k_{-w}=0$
also, and we put
$\rho_w=\pm1=m\ds(w)$.
Since in this case there is no any restriction on the lengths
$l_w$, $l_{-w}$,
we take as
$\la_w$, $\la_{-w}$
in~(\ref{eq:interpr2}) arbitrary positive numbers, and then
(\ref{eq:interpr1}) -- (\ref{eq:interpr3}) are satified with
$\xi_w$, $\xi_{-w}=0$, $\psi_w=0$.
Such edges
$(w,-w)$
separate the graph into several pieces,
and for simplicity we consider the case when there is only one piece.
Moreover, we assume that condition~(\ref{eq:cosnondeg}) with
$k_w$, $k_{-w}>0$
for all
$w\in W_s$
is satisfied for the geometrization of
$M$.
This defines the sign of the function
$\ds$, $\ds(w)>0$
for all
$w\in W_s$.
Take
$w$, $w'\in\dv\cap W_s$
for some vertex
$v\in V$.
It follows from the equality
$l_w=l_{w'}$
that the condition~(\ref{eq:lengthdef}) for
$(\nb,\xi)$
to be critical has the form
$$\frac{k_w^2+k_{-w}^2}{k_w}
  \cdot\frac{2-\rho_w}{\rho_w-1}\cdot\frac{1}{\ds^2(w)}
  =\frac{k_{w'}^2+k_{-w'}^2}{k_{w'}}
   \cdot\frac{2-\rho_{w'}}{\rho_{w'}-1}\cdot\frac{1}{\ds^2(w')},$$
where we used relations~(\ref{eq:interpr1}), (\ref{eq:interpr2}).
Recalling that
$\rho_w=m\ds(w)$,
we obtain

\begin{equation}\label{eq:solmaxw}
 \frac{2-\rho_{w'}}{(\rho_{w'}-1)\rho_{w'}^2}
  =\frac{2-\rho_w}{(\rho_w-1)\rho_w^2}\cdot K_{w,w'},
\end{equation}
where
$$K_{w,w'}=\frac{k_w^2+k_{-w}^2}{k_{w'}^2+k_{-w'}^2}
           \cdot\frac{k_{w'}}{k_w}.$$

Let
$\rho=\rho(\de)$
be the function on
$(0,\infty)$
inverse to the function
$$\de(\rho)=\frac{2-\rho}{(\rho-1)\rho^2}.$$
(It is easy to see that the function
$\de=\de(\rho)$
is monotone on
$1<\rho<2$
and
$\rho(\de)\to 2$
as
$\de\to 0+$, $\rho(\de)\to 1$
as
$\de\to\infty$).

Putting
$\rho_w=\rho(\de)$
for some
$\de>0$,
we obtain that (\ref{eq:solmaxw}) is satisfied for
$\rho_{w'}=\rho(K_{w,w'}\cdot\de)$.
Therefore, values of the function
$\ds$
are also fixed by
$$\ds(w)=\frac{\rho(\de)}{m},\quad\ds(w')=
   \frac{\rho(K_{w,w'}\cdot\de)}{m}.$$

If the graph
$\Ga$
of the manifold
$M$
is simply connected, then the choice
$\ds(w)=\rho(\de)/m$
for an edge
$w\in W_s$
for some
$\de>0$
well and uniquely defines the value
$\ds(w')$
for any edge
$w'\in W_s$
such that condition~(\ref{eq:solmaxw}) is satisfied
for every pair of adjacent edges
$w$, $w'$,
providing that the configuration
$(\nb,\xi)$
is critical.

Assuming that the graph
$\Ga$
is nonsimply connected, let us consider some its nontrivial (oriented) circuit
$Z$.
Defining
$\ds$
as above, we obtain that the function
$\ds$
is well defined, if the condition
$$\prod_{w\in Z}K_{w,w'}=1$$
is satisfied, where
$w'$
is the next edge of
$Z$
after
$w$.
But, obviously, we have
$$\prod_{w\in Z}K_{w,w'}=\prod_{w\in Z}\frac{k_w}{k_{-w}},$$
and the function
$\ds$
is well defined due to condition~(\ref{eq:prod}).
\end{proof}

\vfill\eject

\section{Appendix~A}

We collect here necessary definitions, results and facts
on geometrizations of graph-manifolds from
\cite{BK1}--\cite{BK3}.

In the present work saying about a graph-manifold
$M$
we mean a 3-dimensional closed orientable manifold, for which there exists
a nonempty minimal collection
$E$
of disjoint embedded incompressible tori such that the closure
$M_v$
of each connected component of the complement to
$E$
is homeomorphic to
$F_v\times S^1$,
where
$F_v$
is a compact surface with boundary different from the disc and the annulus.
The manifolds
$M_v$
are called {\it blocks} of
$M$.
The set
$V$
of the blocks is the vertex set of the graph
$\Ga$
of
$M$,
whose set of nonoriented edges is
$E$.
We denote by
$W$
the set of oriented edges of
$\Ga$.

\medskip\noindent
{\bf Waldhausen bases.} We fix an orientation of
$M$.
This defines the orientation of every block
$M_v$,
for which we also fix an orientation of the factor
$S^1$
in the decomposition
$M_v=F_v\times S^1$.

For
$w\in\dv\sub W$
let
$L_w\simeq\Z^2$
be the first homology group of the corresponding boundary component
$(\bd M_v)_w=T_w$,
which is the torus
$T^2$.
Choose a basis
$\set{(z_w,f_w)}{$w\in\dv$}$
of the group
$H_1(\bd M;\Z)=\oplus_{w\in\dv}L_w$
in such a way that the basis
$(z_w,f_w)$
is compatible with the orientation induced on
$\bd M_v$,
the elements
$f_w$
represent oriented fibers, and the sum
$\oplus_{w\in\dv}z_w$
belongs to the kernel of the inclusion homomorphism
$$H_1(\bd M_v;\Z)\to H_1(M_v;\Z).$$
The collection
$(z,f)=\set{(z_w,f_w)}{$w\in\dv,\ v\in V$}$
is called {\it a Waldhausen basis} of
$M$.

To an oriented edge
$w$
of the graph
$\Ga$
there corresponds a gluing map of the boundary components of adjacent blocks,
which induces an isomorphism
$g_w:L_{-w}\to L_w$;
in the chosen bases, the
$g_w$
has the matrix
\[\left[\begin{array}{cc}
           a_w&b_w\\
           c_w&d_w\end{array}\right]\in GL(2,\Z),\]
i.e.
\begin{eqnarray*}
     g_{w}(z_{-w})&=&a_wz_w+c_wf_w\\
     g_{w}(f_{-w})&=&b_wz_w+d_wf_w.
\end{eqnarray*}
We have
$\det g_w=a_wd_w-b_wc_w=-1$
because
$M$
is orientable. Furthermore,
$g_{-w}=g_w^{-1}$
and
$b_{-w}=b_w$.

\medskip\noindent
{\bf Invariants of a graph-manifold.} The (nonoriented)
{\it intersection index}
$b_e=|b_w|=|b_{-w}|$
of an edge
$e=(w,-w)$
is a topological invariant of the manifold
$M$.
We have
$b_e\neq 0$,
since the collection
$E$
is minimal.

{\it The charge}
$$k_v=\sum_{w\in\dv}d_w/b_w$$
of a block
$M_v$
(of a vertex
$v$
of
$\Ga$)
is also an invariant of the oriented manifold
$M$.
The charges change sign if the orientation of
$M$
is changed.

The charges
$k_v$
and the intersection indices
$b_w$
are the coefficients of the compatibility
equation~(\ref{eq:comp}).

\begin{Exa}\label{Exa:dipole1} For a graph-manifold
$M$
whose graph is the dipole there
is only one gluing map
\[g=\left[\begin{array}{cc}
    a&b\\
    c&d\end{array}\right],\quad g^{-1}=
    \left[\begin{array}{cc}
   -d&b\\
    c&-a\end{array}\right].\]

Thus the intersection index of
$M$
is equal to
$|b|$,
and the charges are
$k_0=d/b$, $k_1=-a/b$.
\end{Exa}

\medskip\noindent
{\bf A criterion of geometrizability.} {\it A geometric structure}
of type
$H^2\times\R$
on a block
$M_v$
is a metric for which the universal covering
$\wt M_v$
is isometric to the metric product
$A_v\times\R$,
where
$A_v$
is a convex subset of the hyperbolic plane
$H^2$
bounded by infinite geodesics. Therefore, the boundary
$\bd M_v$
of each geometrized block
$M_v$
is totally geodesic and every its component is a flat torus. Notice that
a geometrized block is not necessarily isometric to the metric product of a
surface
$F_v$
with a hyperbolic metric by a circle. It may happen that there is a
nontrivial holonomy of the circle
$S^1$
along some (noncontractible) loops in the base
$F_v$.
Moreover, such situation is typical and as exception a geometrized block
has a metric splitting globally.

A graph-manifold
$M$
admits {\it a geometrization} if on every its block
$M_v$
a geometric structure of type
$H^2\times\R$
can be chosen in such a way that all gluing maps are isometries.

Each geometrization defines on
$M$
a
$C^1$-smooth
metric, which is real analytic inside of each block. That metric has
nonpositive curvature (the sectional one inside of the blocks and
in the sense of Alexandrov-Busemann everywhere). A graph-manifold
$M$
admits a geometrization iff on
$M$
there exists a
$C^{\infty}$-smooth
Riemannian metric of nonpositive sectional curvature.

The information about geometrizability of
$M$
is encoded in the labeled graph
$\left\{\Ga;B;K\right\}$
of a graph-manifold
$M$.
Here
$B=\set{b_w=b_{-w}\in\N}{$w\in W$}$
is the collection of intersection indices of
$M$,
$K=\set{k_v\in\Q}{$v\in V$}$
is the charge vector. To simplify the statement we restrict to the case
when
$K>0$,
i.e.
$k_v>0$
for all
$v\in V$.

Let
$J_B$
be the quadratic form on
$\R^V$
given by
$$\left(J_B\cL,\cL\right)=
  \sum_{w\in W}\frac{1}{b_w}l_{\dmw}l_{\dpw},$$
where
$\cL=\set{l_v}{$v\in V$}\in\R^V$.

Let $\diag K$ denotes the quadratic form on $\R^V$ given by
$$\left(\diag K\cL,\cL\right)=
   \sum_{v\in V}k_v l_v^2.$$

\begin{Thm}\label{Thm:geometrizability} A graph-manifold
$M$
with positive charge vector
$K$
admits no geometrization if and only if the form
$$H_M=\diag K-J_B$$
is positive semidefinite.
\end{Thm}

In the general case the criterion is similar (see
\cite[Theorem~0.6]{BK2}), only the definition of the form
$H_M$
is more complicated. In other words, this result gives a necessary
and sufficient condition for the compatibility equation~(\ref{eq:comp})
to possess an isometric state.

For instance, if
$k_v=0$
for all
$v\in V$,
then
$M$
admits a geometrization.

Coming back to the case when the graph of
$M$
is the dipole, we obtain
$k_0k_1b^2=-ad$,
and the condition
$0\le k_0k_1b^2<1$
of geometrizability is fulfilled only if
$k_0k_1=0$,
because
$a$, $d\in\Z$.
Then
$k_0=k_1=0$,
i.e.
$a=d=0$
and
$b=c=\pm1$.
Therefore, if such an
$M$
admits a nonpositively curved metric, then the angle between the fibers
$\om=\pi/2$
for any such metric.

\medskip\noindent
{\bf Toral sum and the decomposition principle.} For graph-manifolds
$M'$, $M''$
their toral sum is defined as follows. Let
$U'\subset M_{v'}'$
and
$U''\subset M_{v''}''$
be some saturated regular neighborhoods of fibers of blocks
$M_{v'}'\subset M'$
and
$M_{v''}''\subset M''$.
Gluing the manifolds
$M'\setminus U'$
and
$M''\setminus U''$
by the homeomorphism
$T'\to T''$
of their boundaries induced by some fiber homeomorphism
$U'\to U''$,
we obtain a toral sum
$M=M'+M''$,
which is also a graph-manifold. The result may depend on the choice of
blocks and, also, on the homotopy type of the gluing map. Assume that some
orientations of the manifolds
$M'$, $M''$
are fixed. Then the choice of a Waldhausen basis
$(z',f')$
for
$M'$
is equivalent to the choice of a splitting
$M_{v'}'=F_{v'}'\times S^1$
with oriented factors for every block of
$M'$.
Let
$(z'',f'')$
be a Waldhausen basis for
$M''$.
Then the toral sum
$M=M'+M''$
along the vertices
$v'$, $v''$
that are identified to give the vertex
$v$
is defined by the condition
$M_v=F_v\times S^1$,
where
$F_v$
is the oriented connected sum
$F_{v'}'\#F_{v''}''$
of the surfaces corresponding to the vertices
$v'$, $v''$.
Furthermore, we have
$k_v=k_{v'}+k_{v''}$
for the charges of corresponding blocks of the manifolds
$M$, $M'$, $M''$.
Namely, it is this property on which the decomposition principle from
Sect.~{\bf\ref{sect:decomp}} is based. As an application of that
principle we immediately obtain.

\begin{Exa}\label{Exa:circles} (1) Assume that the graph
$\Ga$
of a manifold
$M$
is a circle with odd number of edges and
$k_v=0$
for all
$v\in V$.
Then for any geometrization of
$M$
all the angles between the fibers of adjacent blocks are
$\om_w=\pi/2$.

This follows from the fact that the labeled graph
$\Ga$
admits only the decomposition into dipoles with zero charges.

(2) Assume that the graph
$\Ga$
of
$M$
is a circle with even number of edges and
$k_v=0$
for all
$v\in V$.
Then
$M$
admits geometrizations with some angles between the fibers of adjacent
blocks
$\om_w$
arbitrary close to zero. This is true for all the angles, if all the
intersection indices are equal to each other.
\end{Exa}

\section{Appendix~B}

Here we derive the Euler-Lagrange equations for the action
$$S(\nb,\xi)=\YM(\nb)+\enn(\xi)-m^2\|\xi\|^2.$$
The computations are done in an invariant form avoiding as
much as possible a choice of coordinates. Thus it is necessary
to deal with
$\cA$-modules
$\hE$, $\hform k$.

We do not assume that the algebra
$\cA$
is commutative. We assume, however, that for the spectral triples
$\{\cA,\cH,ds\}$
we consider the following condition
\begin{equation}\label{eq:commute}
 [ds^2,\pi(a)]=0
\end{equation}
holds for every
$a\in\cA$.
Notice that this condition is fulfilled automatically for the
spectral triples used above.

\subsection{The conjugate module
$\hE$
to
$\cE$}

Let
$\hE$
be the left
$\cA$-module
of
$\cA$
linear homomorphisms
$\cE\to\cA$,
i.e. for
$\ze\in\hE$, $\eta\in\cE$
and
$a\in\cA$
we have
$$\ze(\eta a)=\ze(\eta)a,$$
and the left action of the algebra
$\cA$
is defined by
$(a\ze)(\eta)=a\ze(\eta)$.

There is a standard hermitian structure
$(\eta,\xi)=\xi^*\eta$
on
$\cE$.
Thus one can define the mutually inverse canonical isomophisms
$\lan:\cE\to\hE$, $\lor:\hE\to\cE$,
where
$$(\lan\xi)(\eta)=(\eta,\xi), \quad \lor\ze=e\left(\ze(e)\right)^*,$$
for
$\xi$, $\eta\in\cE$, $\ze\in\hE$.
The definition of
$\lor$
does not depend on the choice of a basis
$e\in\cU$,
since for
$e'=eu^*$
we have
$$e'\left(\ze(e')\right)^*=eu^*\left(\ze(e)u^*\right)^*
    =e\left(\ze(e)\right)^*.$$
It easy follows from definitions that
$\lor\circ\lan=\id_{\cE}$, $\lan\circ\lor=\id_{\hE}$.
Furthermore, for bases
$e\in\cE$, $\ane\in\hE$
we have $\ane(e)=e^*e=1$.

For a section
$\ze\in\hE$
its coordinate
$b=b(\ze,\ane)\in\cA$
with respect to the basis
$\ane$
is defined by the condition
$\ze=b\cdot\ane$.
The gauge group
$\cU$
acts on coordinates of
$\ze$
as follows. If
$e'=eu^*$,
then
$\ane'=u\cdot\ane$,
and for
$\ze=b'\cdot\ane'$
we have
$b'=bu^*$.

For a hermitiam structure
$\La$
on
$\cE$
the corresponding hermitian
$\La$-structure
on
$\hE$
is defined by the condition that the isomorphism
$\lor:\hE\to\cE$
is an isometry of that structures,
$$(\ze,\ka)_{\La}:=(\lor\ka,\lor\ze)_{\La}$$
for
$\ze$, $\ka\in\hE$.
Therefore,
$(\lan\xi,\lan\eta)_{\La}=(\lor\circ\lan\eta,\lor\circ\lan\xi)_{\La}
  =(\eta,\xi)_{\La}$
for
$\xi$, $\eta\in\cE$.
Whereas for
$a$, $b\in\cA$, $\xi$, $\eta\in\cE$
holds
$$(\xi a,\eta b)_{\La}=b^*(\xi,\eta)_{\La}a,$$
for
$\ze$, $\ka\in\hE$
we have
$$(a\ze,b\ka)_{\La}=a(\ze,\ka)_{\La}b^*,$$
because
$\lor(a\ze)=\lor\ze a^*$.

Furthermore, if
$\xi=ea\in\cE$,
then the element
$\lan\xi\in\hE$
has coordinates
$\lan\xi=a^*\ane$.

\subsection{The hermitian metrics}
\subsubsection{The hermitian metric on
$\form k$}

For
$k\ge 0$
and
$\om$, $\om'\in\form k$
we put
$$\lgl\om,\om'\rgl=\tr({\om'}^*\om).$$
It follows from the properties of the trace
$\tr$
that
\begin{eqnarray*}
   \lgl\om',\om\rgl&=&\lgl{\om'}^*,\om^*\rgl
     =\ov{\lgl\om,\om'\rgl};\\
    \lgl a\om,\om'\rgl&=&\lgl\om,a^*\om'\rgl;\\
    \lgl\om b,\om'\rgl&=&\lgl\om,\om'b^*\rgl
\end{eqnarray*}
for
$a$, $b\in\cA$.

\subsubsection{The hermitiam metric on
$\eform k$}

We assume that a hermitian structure
$\La$
on
$\cE$
is fixed. For
$\xi\ot\om$, $\xi'\ot\om'\in\eform k$
we put
$$\lgl\xi\ot\om,\xi'\ot\om'\rgl=
    \lgl\om,(\xi,\xi')_{\La}\om'\rgl,$$
and extend it by linearity on
$\eform k$.
It easy follows from the properties of the hermitian metric on
$\form k$
that
$$\lgl\xi'\ot\om',\xi\ot\om\rgl
   =\ov{\lgl\xi\ot\om,\xi'\ot\om'\rgl}.$$

\subsubsection{The hermitian metric on
$\hform k$}

The space of the endomorphisms
$\End_{\cA}\left(\cE,\eform k\right)$
of right
$\cA$-modules
$\cE$
and
$\eform k$
can be identified with vector space (over
$\C$) $\hform k$
(the tensor products are over
$\cA$).
For
$\xi\ot\om\ot\ze\in\hform k$
and
$\eta\in\cE$
we have
$$(\xi\ot\om\ot\ze)(\eta)=\xi\ot\om\cdot\ze(\eta).$$

We introduce a
$\La$-hermitian metric on
$\hform k$
putting
\begin{eqnarray*}
   \lgl\xi\ot\om\ot\ze,\xi'\ot\om'\ot\ze'\rgl:&=&
    \lgl\xi\ot\om(\ze,\ze')_{\La},\xi'\ot\om'\rgl\\
     &=&\lgl\om(\ze,\ze')_{\La},(\xi,\xi')_{\La}\om'\rgl.
\end{eqnarray*}
In particular, for a basis
$e\in\cU$
we have
$$\lgl e\ot\om\ot\ane,e\ot\om'\ot\ane\rgl
   =\lgl\om(\ane,\ane)_{\La}, (e,e)_{\La}\om'\rgl
   =\lgl\om\la,\la\om'\rgl.$$

\begin{Lem}\label{Lem:hermetric} Assume that an element
$\hat\xi\in\hE$
is conjugated with
$\xi\in\cE$
with respect to the hermitian metrics on
$\eform k$, $\hform k$,
i.e.
$$\lgl\ov\om,\ga(\xi)\rgl=\lgl\ov\om\ot\hat\xi,\ga\rgl$$
for any
$\ov\om\in\eform k$, $\ga\in\hform k$.
Then for a basis
$e\in\cU$
we have
$$\hat a=a^*\la^{-1},$$
where
$\xi=ea$, $\hat\xi=\hat a\cdot\ane$, $\la=\la(\La,e)$.
\end{Lem}

\begin{proof} One can assume that
$\ov\om=e\ot\om$, $\ga=e\ot\si\ot\ane$.
Then
$$\lgl\ov\om,\ga(\xi)\rgl=
   \lgl e\ot\om,e\ot\si\cdot\ane(ea)\rgl=
   \lgl\om,(e,e)_{\La}\si a\rgl=\lgl\om a^*,\la\si\rgl.$$
On the other hand,
\begin{eqnarray*}
  \lgl\ov\om\ot\hat\xi,\ga\rgl&=&
  \lgl e\ot\om\ot\hat a\cdot\ane,e\ot\si\ot\ane\rgl
   =\lgl\om(\hat a\cdot\ane,\ane)_{\La},\la\si\rgl\\
   &=&\lgl\om\hat a\la,\la\si\rgl.
\end{eqnarray*}
Thus
$a^*=\hat a\la$.
\end{proof}

Notice also that the expression
$a^*\la^{-1}\ane$
does not depend on the choice of a basis: for
$e'=eu^*$
we have
$\ane'=u\cdot\ane$, $a'=ua$, ${a'}^*=a^*u^*$, $\la'=u\la u^*$,
${\la'}^{-1}=u\la^{-1}u^*$
and
$${a'}^*{\la'}^{-1}\ane'=a^*u^*u\la^{-1}u^*u\ane=u^*\la^{-1}\ane.$$

\subsection{The Euler-Lagrange equations}

We compute the variation of the action
$S=S(\nb,\xi)$
under a variation
$\nb\mapsto\nb+\dnb$
of a
$\La$-hermitian connection
$\nb$
on
$\cE$
and a variation
$\xi\mapsto\xi+\dxi$
of a section
$\xi\in\cE$,
keeping only linear in
$\dnb$, $\dxi$
terms. We always assume that the connection
$\nb'=\nb+\dnb$
is
$\La$-hermitian.
To state the result we need a number of definitions.

\medskip
(1) For
$\xi\in\cE$
the element
$\hat\xi\in\hE$
is defined by the condition
$$\lgl\om,\ga(\xi)\rgl=\lgl\om\ot\hat\xi,\ga\rgl$$
for all
$\om\in\eform k$, $\ga\in\hform k$,
see Lemma~\ref{Lem:hermetric}.

\medskip
(2) The connection
$\hat\nb:\hE\to\form 1\ot\hE$
on
$\hE$
is defined by
$\nb$
by the condition
$\hat\nb\he=\Phi^*\ot\he$,
where
$\nb e=e\ot\Phi$, $\he=\la^{-1}\cdot\lan e$.
It is easy to see that this definition does not depend on
the choice of
$e\in\cU$.

\medskip
(3) Now we define the current
$J$
as an element of the space
$\hform 1$,
$$J=J(\nb,\xi):=\nb\xi\ot\hat\xi-\xi\ot\hat\nb\hat\xi.$$

\medskip
(4) Let
$\nb^*:\eform 1\to\cE$
be the conjugate to
$\nb$
operator, i.e.
$\lgl\nb^*\om,\eta\rgl=\lgl\om,\nb\eta\rgl$
for every
$\om\in\eform 1$, $\eta\in\cE$.
{\it The generalized Laplacian} is then the operator
$\nb^*\nb:\cE\to\cE$.

\medskip
(5) Let
$d^*:\form *\to\form *$
be the operator conjugated to the exterior differential
$d:\form *\to\form *$, $\lgl d^*\si,\om\rgl=\lgl\si,d\om\rgl$
for all
$\si$, $\om\in\form *$.

The operator
$\cjnb:\hform 2\to\hform 1$
is defined
by
$\cjnb(e\ot\si\ot\he)=e\ot(\om-\om^*)\ot\he$,
where
$$\om=d^*\si+\si\Phi^*-\Phi\si.$$
It will be shown that this definition does not depend on
the choice of
$e\in\cU$.

\begin{Thm}\label{Thm:eleq} The Euler-Lagrange equations
for the action
$S=S(\nb,\xi)$
have the form
\begin{eqnarray}
    \cjnb\nb^2&=&-J\label{eqn:2ndpairapp}\\
  \nb^*\nb\xi&=&m^2\xi.\label{eqn:waveapp}
\end{eqnarray}
\end{Thm}

\subsubsection{The operator
$d^*$}

Before we start the proof of Theorem~\ref{Thm:eleq},
let us first describe the operator
$d^*$
and show that the operator
$\cjnb$
is well defined.

\begin{Lem}\label{Lem:conjdif} For
$\om\in\form 1$
we have
$$d^*\om=-i[D,\om],$$
where recall
$D=ds^{-1}$
is the Dirac operator. In particular,
$(d^*\om)^*=d^*\om^*$.
\end{Lem}

\begin{proof} Let
$a\in\cA$.
Then
\begin{eqnarray*}
   \lgl d^*\om,a\rgl&=&\lgl\om,da\rgl=\tr\left((da)^*\om\right)\\
                   &=&\tr\left(-i(a^*D-Da^*)\om\right)\\
                   &=&-\tr\left(a^*i(D\om-\om D)\right)\\
                   &=&\lgl-i[D,\om],a\rgl.
\end{eqnarray*}
\end{proof}

\begin{Lem}\label{Lem:conjdifform} For
$\om\in\form 1$, $\si\in\form 2$
we have
$$d\om=i(D\om+\om D)\in\form 2;\quad d^*\si=-i(D\si+\si D)\in\form 1.$$
\end{Lem}

\begin{proof} The second equality, obviously, follows from the first one.
In turn, the first equality easily follows from
definition~(\ref{eq:difform}) of the
exterior differential and condition~(\ref{eq:commute}).
\end{proof}

\begin{Cor}\label{Cor:conjdifform} For
$a\in\cA$, $\si\in\form 2$
we have
\begin{eqnarray*}
   d^*(a\si)&=&-da\cdot\si+ad^*\si;\\
   d^*(\si a)&=&d^*\si\cdot a+\si da.
\end{eqnarray*}
\end{Cor}

Let us show now that the operator
$\cjnb$
is well defined. Choosing another basis
$e'=eu^*$,
we have
$\Phi'=udu^*+u\Phi u^*$, $\la'=u\la u^*$, $\hat{e'}=u\he$,
and for
$\si'\in\form2$
with
$e'\ot\si'\ot\hat{e'}=e\ot\si\ot\he$
we obtain
$\si'=u\si u^*$.
Now a straighforward calculation with Corollary~\ref{Cor:conjdifform}
and the relation
$u^*u=uu^*=1$
gives
$$\om'=u\om u^*$$
for
$\om=d^*\si+\si\Phi^*-\Phi\si$
and
$\om'=d^*\si'+\si'{\Phi'}^*-\Phi'\si'$.
This gives the required equality
$$e'\ot(\om'-{\om'}^*)\ot\hat{e'}=e\ot(\om-\om^*)\ot\he.$$

\subsubsection{The energy variation
$\de\enn(\xi)$}

An operator
$\dnb:\cE\to\eform 1$
is
$\cA$-linear
being the difference of two differentiations. Thus it can be considered as
an element of the space
$\hform 1$.
Furthermore, for a basis
$e\in\cU$
we have
$\dnb=e\ot\de\Phi\ot(\lan e)$,
where
$\dnb(e)=e\ot\de\Phi$.

\begin{Lem}\label{Lem:varen} For
$\xi\in\cE$
we have
$$\lgl\dnb,\nb\xi\ot\hat\xi\rgl
  =-\lgl\xi\ot\hat\nb\hat\xi,\dnb\rgl.$$
\end{Lem}

\begin{proof} Since the connections
$\nb$
and
$\nb'=\nb+\dnb$
are
$\La$-hermitian,
the equality
$\la\cdot\de\Phi+(\de\Phi)^*\la=0$
holds. Thus representing
$\xi=ea$,
we have
$\nb\xi=e\ot(da+\Phi a)$, $\hat\xi=a^*\he$
(see Lemma~\ref{Lem:hermetric}) and
\begin{eqnarray*}
   \lgl\dnb,\nb\xi\ot\hat\xi\rgl
   &=&\lgl e\ot\de\Phi\ot(\lan e),e\ot(da+\Phi a)\ot a^*\he\rgl\\
   &=&\lgl\de\Phi,\la(da+\Phi a)a^*\rgl\\
   &=&\lgl a(da+\Phi a)^*\la,(\de\Phi)^*\rgl\\
   &=&\lgl a(da+\Phi a)^*,(\de\Phi)^*\la\rgl\\
   &=&-\lgl a(da+\Phi a)^*,\la\cdot\de\Phi\rgl\\
   &=&-\lgl e\ot a(da+\Phi a)^*\ot\he,e\ot\de\Phi\ot\lan e\rgl\\
   &=&-\lgl\xi\ot\hat\nb\hat\xi,\dnb\rgl.
\end{eqnarray*}
\end{proof}

\begin{Lem}\label{Lem:conjcon} For
$e\ot\om\in\form1$, $\xi=ea\in\cE$
we have
\begin{eqnarray*}
   \nb^*(e\ot\om)
    &=&e\cdot\la^{-1}\left(d^*(\la\om)+\Phi^*\la\om\right);\\
    \nb^*\nb\xi
    &=&e\cdot\la^{-1}\left(d^*\left(\la(da+\Phi a)\right)
      +\Phi^*\la(da+\Phi a)\right),
\end{eqnarray*}
where
$\nb e=e\ot\Phi$.
\end{Lem}

\begin{proof} The second equality follows from the first one, because
$\nb\xi=e\ot(da+\Phi a)$.
To prove the first equality, we take
$\eta\in\cE$, $\eta=e\cdot b$
and consider
\begin{eqnarray*}
  \lgl\nb^*(e\ot\om),\eta\rgl
  &=&\lgl e\ot\om,\nb\eta\rgl
    =\lgl e\ot\om,e\ot(db+\Phi b)\rgl\\
  &=&\lgl\om,\la(db+\Phi b)\rgl
    =\lgl\la\om,db\rgl+\lgl\la\om,\Phi b\rgl\\
  &=&\lgl d^*(\la\om)+\Phi^*\la\om,b\rgl\\
  &=&\lgl\la^{-1}\left(d^*(\la\om)+\Phi^*\la\om\right),
     (e,e)_{\La}b\rgl\\
  &=&\lgl e\cdot\la^{-1}\left(d^*(\la\om)+\Phi^*\la\om\right),\eta\rgl.
\end{eqnarray*}
\end{proof}

For a
$\La$-hermitian
connection
$\nb:\cE\to\eform1$
we define two, in general, different connections
$\hat\nb$, $\ov\nb:\hE\to\form1\ot\hE$
on
$\hE$,
putting for a basis
$e\in\cU$
$$\hat\nb\he=\Phi^*\ot\he\quad\text{and}\quad
  \ov\nb(\lan e)=\Phi^*\ot\lan e,$$
where
$\nb e=e\ot\Phi$, $\he=\la^{-1}\cdot\lan e$.
Since
$\nb$
is
$\La$-hermitian, it follows easily that
$(\hat\nb-\ov\nb)(\lan e)=\la(\Phi+\Phi^*)\la^{-1}\ot\lan e$.
Thus for an unitary connection
$\nb$,
i.e.
$\Phi+\Phi^*=0$,
we have
$\ov\nb=\hat\nb$.

Given
$\xi=e\cdot a\in\cE$,
one has
$\lan\xi=a^*(\lan e)$.
Hence
$\ov\nb(\lan\xi)=(da^*+a^*\Phi^*)\ot\lan e$.
From this, as in Lemma~\ref{Lem:conjcon}, we obtain
(leaving details to the reader)

\begin{Lem}\label{Lem:conjcon1} For
$\om\ot\lan e\in\form1\ot\hE$
and
$\lan\xi=a^*(\lan e)\in\hE$
we have
\begin{eqnarray*}
   \ov\nb^*(\om\ot\lan e)
    &=&\left(d^*(\om\la)+\om\la\Phi\right)\he;\\
   \ov\nb^*\ov\nb(\lan\xi)
    &=&\left(d^*\left((da^*+a^*\Phi^*)\la\right)
    +(da^*+a^*\Phi^*)\la\Phi\right)\he.
\end{eqnarray*}
\end{Lem}

Using Lemmas~\ref{Lem:conjcon} and \ref{Lem:conjcon1},
a straightforward computation gives

\begin{Cor}\label{Cor:conjcon} For
$\xi$, $\dxi\in\cE$
we have
$$\lgl\dxi,\nb^*\nb\xi\rgl=
   \lgl\ov\nb^*\ov\nb(\lan\xi),\lan\dxi\rgl.\qed$$
\end{Cor}

Let
$\de\enn(\xi)$
be the linear in
$\dnb$, $\dxi$
part of the difference
$E_{\nb+\dnb}(\xi+\dxi)-\enn(\xi)$.

\begin{Lem}\label{Lem:envar} The energy variation has the form
$$\de\enn(\xi)=\lgl J,\dnb\rgl
  +\lgl\nb^*\nb\xi,\dxi\rgl
  +\lgl\ov\nb^*\ov\nb(\lan\xi),\lan\dxi\rgl,$$
where
$J=\nb\xi\ot\hat\xi-\xi\ot\hat\nb\hat\xi$.
\end{Lem}

\begin{proof} We have
$$E_{\nb+\dnb}(\xi+\dxi)-\enn(\xi)
  =\lgl(\nb+\dnb)(\xi+\dxi),(\nb+\dnb)(\xi+\dxi)\rgl
   -\lgl\nb\xi,\nb\xi\rgl.$$
Thus using Lemma~\ref{Lem:varen} and Corollary~\ref{Cor:conjcon},
we obtain
\begin{eqnarray*}
 \de\enn(\xi)
  &=&\lgl\nb\xi,\nb(\dxi)\rgl+\lgl\nb(\dxi),\nb\xi\rgl
     +\lgl\nb\xi,\dnb(\xi)\rgl+\lgl\dnb(\xi),\nb\xi\rgl\\
  &=&\lgl\nb^*\nb\xi,\dxi\rgl+\lgl\dxi,\nb^*\nb\xi\rgl
     +\lgl\nb\xi\ot\hat\xi,\dnb\rgl+\lgl\dnb,\nb\xi\ot\hat\xi\rgl\\
  &=&\lgl J,\dnb\rgl+\lgl\nb^*\nb\xi,\dxi\rgl
     +\lgl\ov\nb^*\ov\nb(\lan\xi),\lan\dxi\rgl.
\end{eqnarray*}
\end{proof}

\subsubsection{The variation of the Yang-Mills action
$\YM(\nb)$}

A straightforward computation shows that the operator
$T:=\nb\circ\dnb+\dnb\circ\nb:\cE\to\eform2$
is
$\cA$-linear,
thus it can be considered as an element of the space
$\hform2$.
Furthermore,
$$T=e\ot\left(d(\de\Phi)+\Phi\cdot\de\Phi
      +\de\Phi\cdot\Phi\right)\ot\lan e,$$
where
$\nb e=e\ot\Phi$, $\dnb(e)=e\ot\de\Phi$
for
$\Phi$, $\de\Phi\in\form1$.

\begin{Lem}\label{Lem:topl} Let
$\te\in\form2$
be the curvature operator of the connection
$\nb$, $\nb^2e=e\ot\te$.
Then
$$\lgl\nb^2,T\rgl=\lgl e\ot\om\ot\he,\dnb\rgl,$$
where
$\om=d^*(\te\la)+\te\la\Phi^*-\Phi\te\la$, $\nb e=e\ot\Phi$.
\end{Lem}

\begin{proof} Representing the curvature
$\nb^2$
as
$\nb^2=e\ot\te\ot\lan e$,
we obtain
$$\lgl\nb^2,T\rgl
  =\lgl\te\la,
  \la\left(d(\de\Phi)+\Phi\cdot\de\Phi+\de\Phi\cdot\Phi\right)\rgl.$$
Since the connection
$\nb$
is
$\La$-hermitian,
we have
$$\la\cdot d(\de\Phi)
  =d(\la\cdot\de\Phi)-(\la\Phi+\Phi^*\la)\cdot\de\Phi.$$
Thus
\begin{eqnarray*}
 \lgl\nb^2,T\rgl
   &=&\lgl d^*(\te\la)-\la^{-1}(\la\Phi+\Phi^*\la)^*\te\la,
     \la\cdot\de\Phi\rgl\\
   &\ &+\lgl\Phi^*\la\te\la,\de\Phi\rgl
      +\lgl\te\la\Phi^*,\la\cdot\de\Phi\rgl\\
   &=&\lgl d^*(\te\la)-\la^{-1}(\la\Phi+\Phi^*\la)\te\la
     +\la^{-1}\Phi^*\la\te\la+\te\la\Phi^*,\la\cdot\de\Phi\rgl\\
   &=&\lgl\om,\la\cdot\de\Phi\rgl
    =\lgl e\ot\om\ot\he,\dnb\rgl.
\end{eqnarray*}
\end{proof}

Using as in the proof of Lemma~\ref{Lem:varen} the equality
$\la\cdot\de\Phi+(\de\Phi)^*\la=0$,
a similar computation gives (we leave details to the reader)

\begin{Lem}\label{Lem:topr} Under the conditions of
Lemma~\ref{Lem:topl} we have
$$\lgl T,\nb^2\rgl=-\lgl e\ot\om^*\ot\he,\dnb\rgl.\qed $$
\end{Lem}

Let
$\de\YM(\nb)$
be the linear in
$\dnb$
part of the difference
$\YM(\nb+\dnb)-\YM(\nb)$.

\begin{Lem}\label{Lem:ymvar} For the variation of
the Yang-Mills action we have
$$\de\YM(\nb)=\lgl\cjnb\nb^2,\dnb\rgl.$$
\end{Lem}

\begin{proof} Since
$\nb^2=e\ot\te\ot\lan e=e\ot\te\la\ot\he$,
applying Lemmas~\ref{Lem:topl}, \ref{Lem:topr}, we obtain
\begin{eqnarray*}
  \de\YM(\nb)
    &=&\lgl\nb^2,T\rgl+\lgl T,\nb^2\rgl\\
    &=&\lgl e\ot(\om-\om^*)\ot\he,\dnb\rgl\\
    &=&\lgl\cjnb\nb^2,\dnb\rgl,
\end{eqnarray*}
where
$\om=d^*(\te\la)+\te\la\Phi^*-\Phi\te\la$.
\end{proof}

\begin{Cor}\label{Cor:actvar} For the variation of the action
$S$
we have
\begin{eqnarray*}
  \de S(\nb,\xi)
   &:=&\de\YM(\nb)+\de\enn(\xi)-m^2\de\|\xi\|^2\\
   &=&\lgl\cjnb\nb^2+J,\dnb\rgl\\
   &\ &+\lgl\nb^*\nb\xi-m^2\xi,\dxi\rgl\\
   &\ &+\lgl\ov\nb^*\ov\nb\lan\xi-m^2\lan\xi,\lan\dxi\rgl.
\end{eqnarray*}
\end{Cor}

\begin{proof} This follows from Lemmas~\ref{Lem:envar},
\ref{Lem:ymvar} and that
$$\de\|\xi\|^2
   =\lgl\xi,\dxi\rgl+\lgl\dxi,\xi\rgl
   =\lgl\xi,\dxi\rgl+\lgl\lan\xi,\lan\dxi\rgl.$$
\end{proof}

\subsubsection{Proof of Theorem~\ref{Thm:eleq}}
Since for
$\dxi=e\cdot\de a$, $\lan\dxi=(\de a)^*\lan e$
we have
$\de a=\re(\de a)+i\cdot\im(\de a)$,
$(\de a)^*=\re(\de a)-i\cdot\im(\de a)$,
one can assume that the variations
$\dxi$, $\lan\dxi$
are independent. Thus the Euler-Lagrange equations for the action
$S$
have the form
\begin{eqnarray*}
  \cjnb\nb^2&=&-J;\\
  \nb^*\nb\xi&=&m^2\xi;\\
  \ov\nb^*\ov\nb\lan\xi&=&m^2\lan\xi.
\end{eqnarray*}
Using that for
$\om\in\form1$
one has
$(d^*\om)^*=d^*\om^*$
(see Lemma~\ref{Lem:conjdif}),
we obtain from Lemmas~\ref{Lem:conjcon} and \ref{Lem:conjcon1}
that the third equation is
implied by the second one.
\qed

Choosing a basis
$e\in\cU$
and representing
$\nb e=e\ot\Phi$, $\xi=ea$, $\la=\la(\La,e)$,
$\hat\xi=a^*\he$,
we obtain
\begin{eqnarray*}
 J&=&\nb\xi\ot\hat\xi-\xi\ot\hat\nb\hat\xi\\
  &=&e\ot\left((da+\Phi a)a^*-a(da+\Phi a)^*\right)\ot\he.
\end{eqnarray*}
Hence, it follows from Lemmas~\ref{Lem:conjcon}, \ref{Lem:ymvar}
that the Euler-Lagrange equations~(\ref{eqn:2ndpairapp}),
(\ref{eqn:waveapp})
in the coordinates have the form
\begin{equation}\label{eq:2ndpaircoor}
   \om-\om^*=a(da+\Phi a)^*-(da+\Phi a)a^*
\end{equation}
\begin{equation}\label{eq:wavecoor}
   d^*\left(\la(da+\Phi a)\right)
    +\Phi^*\la(da+\Phi a)=m^2\la a,
\end{equation}
where
$\om=d^*(\te\la)+\te\la\Phi^*-\Phi\te\la$.

For the dipole case we let
$p=(1,0)$, $q=1-p\in\cA$
be projectors. Then
$a=\xi_0p+\xi_1q$, $\la=\la_0p+\la_1q$,
$\Phi=-i\ds^2(\phi_{01}pdp+\phi_{10}qdp)$,
$\te=i(\phi_{10}-\phi_{01}+i\phi_{10}\phi_{01}\ds^2)\cdot 1$,
where
$\xi_0$, $\xi_1$, $\phi_{01}$, $\phi_{10}\in\C$,
$\la_0$, $\la_1>0$.
Straightforward calculations with using the
$\La$-hermitian
condition~(\ref{eq:newcompcon}) and Lemmas~\ref{Lem:conjdif},
\ref{Lem:conjdifform} for finding of the exterior
differential
$d^*$
show that the equation~(\ref{eq:2ndpaircoor}) takes the form
(\ref{eq:twopoint2nd}) and (\ref{eq:wavecoor})
the form (\ref{eqn:twopointmaxwell}). One should use on some step
that the operator
$\te$
is selfadjoint due to condition~(\ref{eq:newcompcon}),
$\te^*=\te$.


\bigskip

\noindent
St.-Petersburg Dept. of Steklov Math. Institute

\noindent
Fontanka 27,

\noindent
191011, St.-Petersburg, Russia

\noindent
{\tt buyalo@pdmi.ras.ru}

\end{document}